\documentclass[aps,prb,superscriptaddress,noshowpacs,noshowkeys,twocolumn,reprint,amssymb,amsmath,amsfonts]{revtex4-1}

\usepackage{graphicx}
\usepackage{theorem}
\usepackage{dcolumn}
\usepackage{longtable}
\usepackage{bm}
\usepackage{accents}
\usepackage{color}

\newcommand{\ie}{{\textit{i.e.}}}
\newcommand{\eg}{{\textit{e.g.}}}

\newcommand{\fp}{f_{\text{p}}}
\newcommand{\fr}{f_{\text{r}}}
\newcommand{\fq}{f_{\text{q}}}

\newcommand{\fs}{f_{\text{s}}}
\newcommand{\Vt}{V_{\text{t}}}
\newcommand{\Vrg}{V_{\text{rg}}}
\newcommand{\Vr}{V_{\text{r}}}
\newcommand{\Vtg}{V_{\text{tg}}}
\newcommand{\Vss}{V_{\text{ss}}}

\renewcommand{~}{\,}

\makeatletter
\renewcommand{\figurename}{Fig.}
\renewcommand{\tablename}{Tab.}
\renewcommand{\fnum@figure}[1]{\textbf{\figurename~\thefigure~\textbar} }
\renewcommand{\fnum@table}[1]{\textbf{\tablename~\thetable~\textbar} }
\makeatother

\begin{document}


\title{Single electrons on solid neon as a solid-state qubit platform}

\author{Xianjing Zhou}
\affiliation{Center for Nanoscale Materials, Argonne National Laboratory, Lemont, Illinois 60439, USA}

\author{Gerwin Koolstra}
\affiliation{Computational Research Division, Lawrence Berkeley National Laboratory, Berkeley, California 94720, USA}

\author{Xufeng Zhang}
\affiliation{Center for Nanoscale Materials, Argonne National Laboratory, Lemont, Illinois 60439, USA}

\author{Ge Yang}
\affiliation{The NSF AI Institute for Artificial Intelligence and Fundamental Interactions, USA}
\affiliation{Computer Science and Artificial Intelligence Laboratory, Massachusetts Institute of Technology, Cambridge, Massachusetts 02139, USA}

\author{Xu Han}
\affiliation{Center for Nanoscale Materials, Argonne National Laboratory, Lemont, Illinois 60439, USA}

\author{Brennan Dizdar}
\affiliation{James Franck Institute and Department of Physics, University of Chicago, Chicago, Illinois 60637, USA}

\author{Xinhao Li}
\author{Ralu Divan}
\affiliation{Center for Nanoscale Materials, Argonne National Laboratory, Lemont, Illinois 60439, USA}

\author{Wei Guo}
\affiliation{National High Magnetic Field Laboratory, Tallahassee, Florida 32310, USA}
\affiliation{Department of Mechanical Engineering, FAMU-FSU College of Engineering, Florida State University, Tallahassee, Florida 32310, USA}

\author{Kater W. Murch}\email{murch@physics.wustl.edu}
\affiliation{Department of Physics, Washington University in St. Louis, St. Louis, Missouri 63130, USA}

\author{David I. Schuster}
\affiliation{James Franck Institute and Department of Physics, University of Chicago, Chicago, Illinois 60637, USA}
\affiliation{Pritzker School of Molecular Engineering, University of Chicago, Chicago, Illinois 60637, USA}

\author{Dafei Jin}\email{djin@anl.gov}
\affiliation{Center for Nanoscale Materials, Argonne National Laboratory, Lemont, Illinois 60439, USA}

\date{\today}

\begin{abstract}

\end{abstract}

\maketitle
\pretolerance=9000 

\textbf{Progress toward the realization of quantum computers requires persistent advances in their constituent building blocks -- qubits. Novel qubit platforms that simultaneously embody long coherence, fast operation, and large scalability offer compelling advantages in the construction of quantum computers and many other quantum information systems~\cite{Ladd2010,Popkin2016,DeLeon2021}. Electrons, ubiquitous elementary particles of nonzero charge, spin, and mass, have commonly been perceived as paradigmatic local quantum information carriers. Despite superior controllability and configurability, their practical performance as qubits via either motional or spin states depends critically on their material environment~\cite{Hanson2007,Zwanenburg2013,DeLeon2021}. Here we report our experimental realization of a new qubit platform based upon isolated single electrons trapped on an ultraclean solid neon surface in vacuum~\cite{Cole1969,Cole1971,leiderer1992electrons,Platzman1999,smolyaninov2001electrons,Dykman2003,Lyon2006,Bradbury2011}. By integrating an electron trap in a circuit quantum electrodynamics architecture~\cite{Wallraff2004,Blais2021,Schuster2010,Yang2016,Koolstra2019,jin2020quantum,Clerk2020}, we achieve strong coupling between the motional states of a single electron and a single microwave photon in an on-chip superconducting resonator. Qubit gate operations and dispersive readout are implemented to measure the energy relaxation time $T_1$ of $15~\mu$s and phase coherence time $T_2$ over $200$~ns. These results indicate that the electron-on-solid-neon qubit already performs near the state of the art as a charge qubit~\cite{Chatterjee2021}.}

The rapid growth of quantum information science and technology in recent years accompanies the remarkable success of various qubit platforms in various domains of quantum information processing. Notable examples include superconducting quantum circuits~\cite{nakamura1999coherent,Wallraff2004,Schoelkopf2008,Clarke2008,Arute2019,Blais2021}, semiconductor quantum dots~\cite{Mi2017,Mi2018,Samkharadze2018,Landig2018,petit2020universal,Burkard2020}, electromagnetically trapped ions~\cite{monroe1995demonstration,kielpinski2002,leibfried2003quantum,Bruzewicz2019,Pino2021}, optically trapped atoms~\cite{brennen1999quantum,jaksch2000fast,Saffman2010,wang2016single}, natural or implanted defects~\cite{Pla2012, Pla2013, Chen2020, Wolfowicz2021}, and magnetic molecules~\cite{Vincent2012,Thiele2014,Atzori2019,Coronado2020}. Among different quantum information carriers, isolated single electrons~---~paradigmatic charged spin-$\frac{1}{2}$ massive particles that naturally interact with photons via quantum electrodynamics (QED)~---~offer conceivably the straightest approach for efficient manipulation and remote entanglement. So far, electron qubits have been made predominantly in semiconductor heterojunctions and semiconductor-oxide interfaces~\cite{DeLeon2021,Hanson2007,Zwanenburg2013}. Despite standardized device fabrication and convenient electrical control, a major challenge faced by these electron qubits is the limited coherent time due to material imperfections or background noise~\cite{DeLeon2021,Hanson2007,Zwanenburg2013}. In this circumstance, a new type of single-electron qubit, embedded in an ultraclean low-noise environment, may open up unprecedented opportunities to resolve the coherence challenge. Along with the inherent features of fast operation and large scalability, this single-electron qubit platform holds great potential for development into an ideal quantum computing architecture in the future.

\begin{figure*}[hbt]
	\centerline{\includegraphics[scale=1]{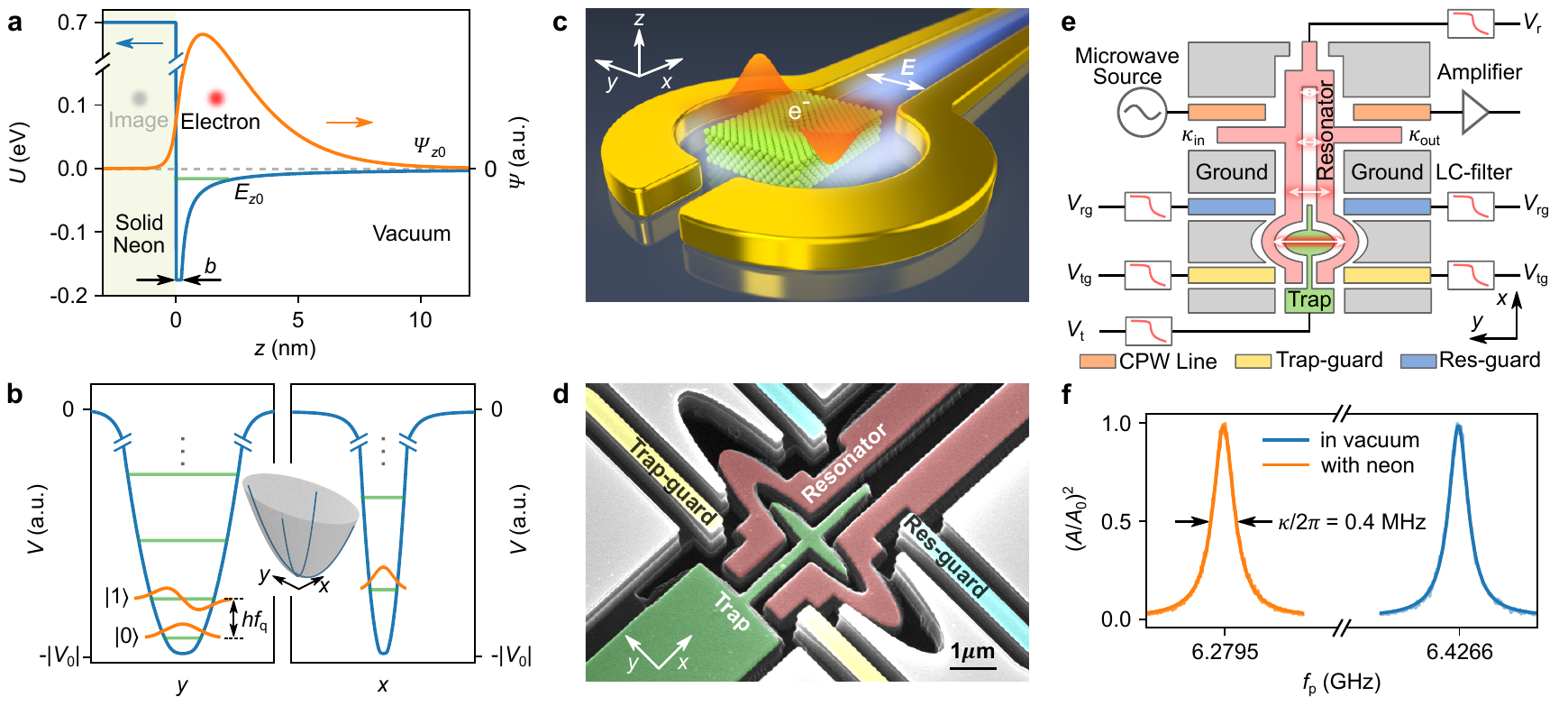}}
	\caption{\textbf{Electronic structure and device design of the single-electron circuit quantum electrodynamics architecture on solid neon}. \textbf{a}, Potential energy seen by an excess electron approaching a flat solid neon surface and calculated ground-state eigen-energy and wavefunction in the $z$-motion. \textbf{b}, Schematic of in-plane trapping potential that defines the motional state qubit in $y$ direction. \textbf{c}, Conceptual illustration of a single electron trapped on solid neon surface and interacting with microwave photons at the open end of a superconducting coplanar stripline resonator. The electric dipole transition and the electric field of microwave photons are aligned in $y$ direction. \textbf{d}, Scanning electron microscopy (SEM) picture of the fabricated device around the electron trap and photon coupling region. The trap and two striplines reside inside a 3.5~$\mu$m wide and 1.5~$\mu$m deep etched channel. \textbf{e}, Specific device structure and functionalities of different components. The superconducting quarter-wavelength double-stripline resonator is measured in transmission by the input and output through coplanar waveguides (CPWs) with coupling rates $\kappa_{\mathrm{in}}$ and $\kappa_{\mathrm{out}}$ respectively. The antisymmetric mode has its electric field maximum at the open end and field direction as indicated by the arrows. The resonator is biased with a DC voltage $V_{\rm{r}}$ to control the number of electrons in the reservoir. Three additional DC electrodes (trap, trap-guard, res-guard), biased with the voltages $\Vt$, $\Vtg$, $\Vrg$, control the single-electron trapping and frequency detuning. Each electrode has its own on-chip low-pass LC filter. \textbf{f}, Measured resonance peaks before ($\fr=6.4266$~GHz) and after ($\fr=6.2795$~GHz) neon fully fills the channel, showing the fitted resonator linewidth $\kappa/2\pi=0.4$~MHz. } \label{Fig:Structure}
\end{figure*}

In this work, we demonstrate a fundamentally new solid-state single-electron qubit platform based upon trapping and manipulating isolated single electrons on an ultraclean solid neon surface in vacuum. By integrating the electron trap in a hybrid circuit QED architecture~\cite{Wallraff2004,Blais2021,Schuster2010,Yang2016,Koolstra2019,jin2020quantum,Clerk2020}, we observe vacuum Rabi splitting between the motional states of a single electron and a single microwave photon in an on-chip superconducting resonator. This observation lays the foundation for the quantum coherent control and (single-shot) dispersive readout of electron charge (motional-state) qubits at microwave frequencies in this system. By detuning the electron transition frequency with respect to the resonator frequency, we perform complete qubit characterization, \ie, two-tone qubit spectroscopy~\cite{Schuster2005} and time-domain measurements~\cite{Wallraff2005}, including Rabi oscillations, $T_1$ energy relaxation time, and $T_2$ phase coherence time measurements. Without optimization, the measured $T_1 = 15~\mu$s and $T_2 \gtrsim 200~$ns have already reached the state of the art for a charge qubit~\cite{Chatterjee2021}, highlighting the promise of this new material environment. With projected development employing spin-charge conversion~\cite{Schuster2010,Mi2018}, we anticipate the nearly perfect spinless environment formed by solid neon (the naturally occurring 0.27\% abundance of spinful $^{21}$Ne can be easily purified away) to support electron spin qubits with estimated coherence time over 1~s~\cite{Lyon2006,Schuster2010,Sheludiakov2019,jin2020quantum}. Beyond quantum computing, this novel solid-state single-electron qubit platform creates an appealing hybrid quantum framework that can connect various qubit platforms, thereby paving new pathways in quantum information science and technology.

\noindent\textbf{Electronic structure and device design}

Neon (Ne) is a noble-gas element next to helium (He) in the periodic table. In contrast to He, which is a liquid (superfluid) even at zero temperature, unless a large pressure of at least 25~bar is applied, Ne spontaneously turns into a face-centered-cubic (fcc) crystal after passing its triple point at the elevated temperature $T_{\text{t}}=24.556$~K and moderate pressure $P_{\text{t}}=0.43$~bar~\cite{jacobsen1997thermodynamic, Pollack1964, Batchelder1967}. At near-zero temperature, solid Ne can form a free surface to vacuum and serve as an ultraclean substrate with no uncontrollable impurities or electromagnetic noise~\cite{zavyalov2005electron,leiderer2016cryocrystals}.

\begin{figure*}[htb]
	\centerline{\includegraphics[scale=1]{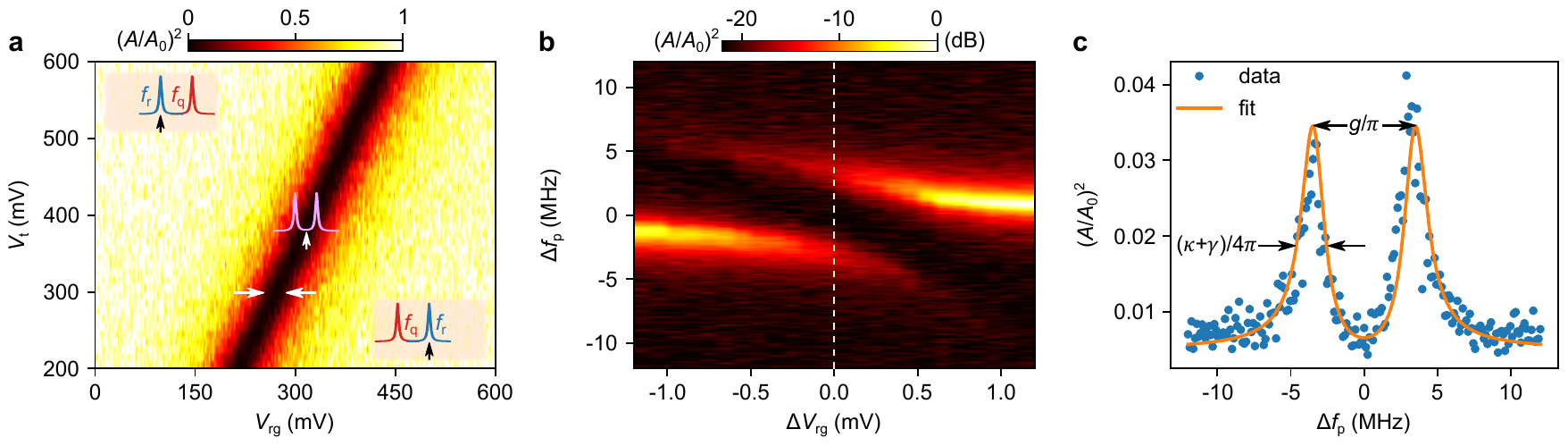}}
	\caption{\textbf{Strong coupling and vacuum Rabi splitting between a single electron on solid neon and a single microwave photon in a superconducting resonator}. \textbf{a}, Normalized microwave transmission amplitude ($A/A_0)^2$ probed at the resonator frequency $\fr$  (black and white vertical arrows in the insets) versus the trap voltage $\Vt$ and the resonator-guard voltage $\Vrg$. The amplitude drops when the electron's transition frequency $\fq$ is tuned on resonance with the resonator. \textbf{b}, Normalized $(A/A_0)^2$ versus the detuned probe frequency $\Delta f_{\text{p}} = f_{\text{p}}-f_{\text{r}}$ and resonator-guard voltage $\Delta \Vrg$ detuned from the resonance condition in the region indicated by the white horizontal arrows in (\textbf{a}) with fixed $\Vt$. \textbf{c},  Line cut from (\textbf{b}) along the white dashed line when the qubit frequency $\fq$ is on resonance with the resonator frequency $\fr$. The two peaks show the vacuum Rabi splitting, giving the coupling strength $g/2\pi=3.5$~MHz and intrinsic electron linewidth $\gamma/2\pi=1.7$~MHz, fitted by input-output theory with the known resonator linewidth $\kappa/2\pi=0.4$~MHz.}
	\label{Fig:Coupling}
\end{figure*}

When an excess electron approaches a semi-infinite solid Ne surface at $z=0$ from vacuum, two effects lead to an out-of-plane trapping potential that can bind the electron to the surface (see Fig.~\ref{Fig:Structure}a). A repulsive barrier, $U\approx0.7$~eV, occurs due to the Pauli exclusion between the excess electron and atomic shell electrons. In addition, an attractive polarization potential, $V(z)= -(\epsilon-1)/ [(\epsilon+1) e^2/4z], (z>b)$, with a dielectric constant $\epsilon = 1.244$ and short-range cutoff $b\approx 2.3$~{\AA}, occurs due to the induced image charge inside solid Ne~\cite{Cole1969,Cole1971,jin2020quantum}.  With this potential, the electron's $z$-motion has a ground-state energy $E_{z0}=-15.8$~meV and an eigen-wavefunction peaked at about 1~nm distance from the surface (Fig.~\ref{Fig:Structure}a).  The energy cost to bring the electron to the first excited state in $z$ is 12.7~meV, equivalent to a 147~K activation temperature. Therefore, at our $10$~mK experimental temperature, the electron is frozen within the ground-state subband of $z$-motion. Previous studies have verified that solid Ne surface can hold a nondegenerate two-dimensional electron gas with $\sim 10^{10}$~cm$^{-2}$ high density and $\sim 10^4$~cm$^2$~V$^{-1}$~s$^{-1}$ high mobility~\cite{Kajita1984}.

Condensed (liquid or solid) noble-gas elements with positive (repulsive) electron affinity are the only materials in nature that can hold electrons on a free surface in vacuum. Practically, all other materials, even electronically insulating and atomically smooth, have negative (attractive) electron affinity and contain charged contaminants or dangling bonds on the surface that can capture and localize excess electrons at atomic to molecular scales~\cite{nilsson2011chemical,ibach2006physics}. While the electron-on-solid-Ne (eNe) system can be considered conceptually as an extension to the historically more studied electron-on-liquid-He (eHe) system, it exhibits much stronger surface rigidity that suppresses decoherence through surface excitations~\cite{Schuster2010,Yang2016,Koolstra2019}. Compared with eHe that was proposed as a qubit platform over two decades ago~\cite{Platzman1999,Dykman2003,Lyon2006,Schuster2010,Bradbury2011,Yang2016,Koolstra2019}, eNe embodies a potentially transformative solid-state qubit platform~\cite{smolyaninov2001electrons,zavyalov2005electron,jin2020quantum}.

On a flat Ne surface, the electron takes plane-wave eigenstates in the $xy$ plane. To confine the electron in the plane, we utilize carefully designed lateral trapping electrodes to hold the electrons individually and deterministically in space~\cite{Koolstra2019}, with trapping time exceeding two months. We tune the electrode voltages to further constrain the electron's $x$-motion to its ground state and take the two lowest energy states of $y$-motion as the qubit states (see Fig.~\ref{Fig:Structure}b). Figure~\ref{Fig:Structure}c shows a simplified conceptual illustration of our arrangement of the trapped electron and microwave resonator. The electron is on the solid Ne surface at the open end (in a ``clamp" shape) of a quarter-wavelength coplanar double-stripline resonator, which carries a symmetric and an antisymmetric mode~\cite{Koolstra2019, Pozar2011}. Both modes have the electric field maximum at the open end, but only the antisymmetric mode has the field direction aligned with electron's motion in $y$. Figure~\ref{Fig:Structure}d displays a scanning electron microscopy (SEM) image of the actual device structure around the trapping area. All the metal lines and ground planes are made of superconducting niobium (Nb) deposited on a high-resistivity silicon substrate. A ``trap" electrode, applied with positive voltage, plugs into the open end of the ``resonator". Four ``guard" electrodes, named as ``trap guards" and ``res(onator) guards" surrounding the trap, applied with voltages in pairs, provide precise tuning to the trapping potential and thus the electron transition frequency about the resonator frequency. The trap and resonator reside inside a $\sim4$~mm long etched channel.

\begin{figure*}[htb]
	\centerline{\includegraphics[scale=1]{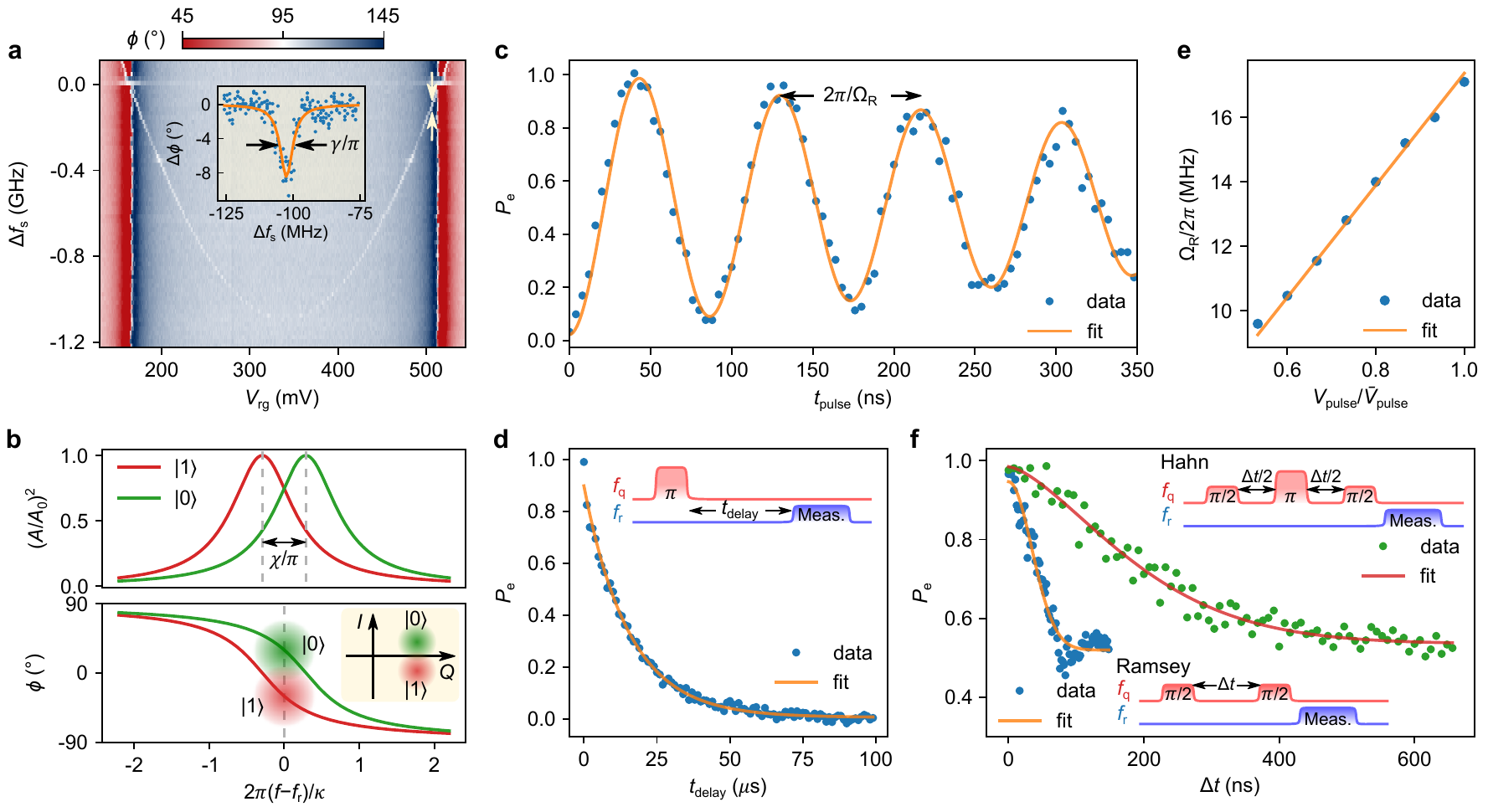}}
	\caption{\textbf{Spectroscopy and time-domain characterization of a single-electron qubit on solid neon}. \textbf{a}, Two-tone qubit spectroscopy measurement on the transmission phase $\phi$ at the resonator frequency $\fr$ versus the detuned pump frequency $\Delta f_{\text{s}}=f_{\text{s}} - \fr$ and the resonator-guard voltage $V_{\rm rg}$. The qubit linewidth $\gamma$ can be obtained by fitting the phase dip profile (inset). \textbf{b}, Illustration of the dispersive readout of single-electron qubit state by transmission measurement. Ground and excited electron states cause the actual resonator frequency to be blueshifted and redshifted and the measured transmission phase $\phi$ at the bare resonator frequency $\fr$ to show on average $\pm30^\circ$ shift. \textbf{c}, Rabi oscillations of the excited state population $P_{\mathrm{e}}$, measured at fixed pulse amplitude and varied pulse length $t_{\mathrm{pulse}}$. \textbf{d}, Qubit relaxation measurement with the fitted relaxation time $T_1=15~\mu$s. \textbf{e}, Measured linear dependence of Rabi frequency $\Omega_{\mathrm{R}}$ on the amplitude of gate pulses. \textbf{f}, Ramsey fringe and Hahn echo measurements with the fitted original coherence time $T_2^*=50$~ns and extended coherence time $T_{\text{2E}}=220$~ns. Qubit detuning for (\textbf{b}) -- (\textbf{f}) is kept at $\Delta/2\pi \equiv f_{\text{q}}-f_{\text{r}}=-100$~MHz. } \label{Fig:Characterization}
\end{figure*}

Figure~\ref{Fig:Structure}e details the structure of this hybrid circuit QED device. The double stripline resonator is coupled with coplanar waveguides (CPWs) with the input and output coupling rates $\kappa_{\mathrm{in}}$ and $\kappa_{\mathrm{out}}$, respectively, in a transmission measurement configuration. Each DC electrode, biased at $\Vr$, $\Vt$, $\Vrg$, $\Vtg$, respectively, has its own on-chip low-pass LC filter that isolates the electron and resonator from the DC electrodes at microwave frequencies to protect the qubit lifetime and resonator quality factor. We fill liquid Ne into the experimental cell at 26~K and cool down to 10~mK. The shift of resonator frequency $\fr$ can be used to infer Ne thickness. When the channel is fully filled with Ne, $\fr$ shifts from 6.4266~GHz to 6.2795~GHz (see Fig.~\ref{Fig:Structure}f). In practice, we only put in a tiny amount of Ne to coat the device surface, resulting only 0.3--0.6~MHz frequency shift and 5--10~nm estimated Ne thickness from numerical simulation. The observed resonator linewidth $\kappa/2\pi=0.4$~MHz, independent of Ne filling, indicates a quality factor $Q\approx1.6\times10^4$. Electrons are generated through thermionic emission from a pair of tungsten filaments inside the cell under a voltage pulse train (width: 0.1~ms, height: 4~V, and repetition rate: 1~kHz) for a total duration of 1~s~\cite{Yang2016,Koolstra2019}. (See Methods for more details.)

\noindent\textbf{Strong coupling and vacuum Rabi splitting}

We use a similar scheme in our previous work~\cite{Koolstra2019} to load single electrons onto the trap from the channel hosting the stripline resonator, and in the meanwhile, monitor the microwave transmission signal. Once an electron is trapped, we fix the resonator voltage $\Vr$ at 1~V and trap-guard voltage $\Vtg$ at 0~V. A positive $\Vr$ is necessary to keep any remnant electrons inside the long channel far off resonance. The trap voltage $\Vt$ and resonator-guard voltage $V_{\rm rg}$ are enough to tune the qubit frequency $\fq$ into resonance with the resonator frequency $\fr \approx 6.426$~GHz. Figure~\ref{Fig:Coupling}a gives a color plot of the normalized transmission amplitude $(A/A_0)^2$ probed at the resonator frequency $\fr$ versus the tuning voltages $\Vt$ and $\Vrg$ for a trapped electron. When $\fq$ is tuned across $\fr$, we observe a sharp drop in the microwave transmission amplitude probed at $\fr$~\cite{Wallraff2004}. The average photon occupancy in the resonator is controlled at the single-photon level, $\bar{n}\approx 1$, with about $-135$~dBm input power to the resonator. (See Methods for more details.)

By measuring the transmission spectrum as a function of probe frequency $\fp$ near the resonator frequency $\fr$, we observe a clear avoided crossing -- vacuum Rabi splitting, when $\Vt$ is fixed and $\Vrg$ tunes the qubit frequency across the resonator frequency (see Fig.~\ref{Fig:Coupling}b). Figure~\ref{Fig:Coupling}c shows the line cut of Fig.~\ref{Fig:Coupling}b in the on-resonance case $\fq=\fr$. A fit over the two peaks by input-output theory~\cite{walls2007quantum} yields the coupling strength $g/2\pi= 3.5$~MHz, which is nearly twice of the electron dephasing rate $\gamma/2\pi\approx1.7$~MHz, with the known resonator decay rate $\kappa/2\pi\approx 0.4$~MHz. (See Methods for more details.) The system has clearly entered the strong coupling regime~\cite{Wallraff2004}, $g>\gamma, \kappa$, which instantly enables coherent microwave control and dispersive readout of single-electron qubits in this system.

\noindent\textbf{Spectroscopy and time-domain characterization}

Figure~\ref{Fig:Characterization}a shows a two-tone qubit spectroscopy measurement of another trapped electron, which has a stronger coupling strength and a wider electron linewidth than the previous one. The qubit frequency $\fq$ is tuned by the resonator-guard voltage $\Vrg$. At each given $\Vrg$, we monitor the transmission phase $\phi$ at the bare resonator frequency $\fr$ while a pump-tone frequency $\fs$ is slowly swept over a range of  $\sim1$~GHz across the qubit frequency $\fq$. When $\fs$ is resonant with $\fq$, it partially excites the qubit, inducing a dip ($\fq < \fr$) or a peak ($\fq > \fr$) in the $\phi$ versus $\fs$ plot (Fig.~\ref{Fig:Characterization}a inset). By scanning both $\fs$ and $\Vrg$, we obtain the intrinsic qubit spectrum as a function of $\Vrg$. A Lorentzian fit of this qubit spectrum yields the linewidth $\gamma/2\pi = 2.8$~MHz. The pump-tone power is kept low enough here to avoid power broadening of the qubit over its natural linewidth. The overall spectrum resembles that of a double-quantum-dot qubit (DQD) spectrum at a semiconductor interface~\cite{Mi2017}. (See Methods for more discussion.)

By increasing the pump-tone power, we investigate the anharmonicity under a varied detuning $\Delta/2\pi \equiv \fq - \fr$ between $\pm100$~MHz. At $-100$~MHz detuning, the measured anharmonicity between the two lowest transition frequencies is $\alpha/2\pi \equiv f_{|1\rangle\rightarrow|2\rangle} - f_{|0\rangle\rightarrow|1\rangle} \approx 40$~MHz, where $|0\rangle$, $|1\rangle$, $|2\rangle$ are the ground, first-excited, and second-excited states, respectively (see Methods). This $\alpha$ is positive and consistent with the value expected from the trap design~\cite{Koolstra2019}. Even though this $\alpha$ value currently limits single qubit gate durations to $t_\pi \gg \pi/\alpha\approx 12$~ns, it can be easily enhanced in future trap designs.

The measured coupling strength for this electron is $g/2\pi=4.5$~MHz (see Methods). At large detuning $\Delta/2\pi = -100$~MHz, we have $|\Delta|\gg g $. The dispersive coupling between the qubit and resonator provides a qubit-state-dependent frequency shift $\chi$. By measuring the transmission phase $\phi$ of the photons at $\fr$, we can read out the qubit state~\cite{Schuster2005,DavidIsaacSchuster2007}. Figure~\ref{Fig:Characterization}b illustrates the scheme of dispersive readout in accordance with our experiment, as is common in circuit QED platforms~\cite{Wallraff2005}. The measured $\phi$ shift at $\fr$ is about $\pm 30^\circ$ after statistical average, corresponding to $\chi/2\pi \approx 0.12$~MHz. Ideally, the qubit would operate at the spectrum minimum of Fig.~\ref{Fig:Characterization}a, \ie, the ``sweet spot" where the charge noise has the lowest effect. However, for this particular electron at the ``sweet spot", the $>1$~GHz large detuning precludes state readout with reasonable signal-to-noise ratio.

We now use real-time coherent control to measure the coherence properties of the qubit at $\Delta/2\pi = -100$~MHz detuning. Figure~\ref{Fig:Characterization}c displays Rabi oscillations in the excited-state population $P_\text{e}$~\cite{Krantz2019}. Starting with the qubit in its ground state, we apply a pulse with variable duration and fixed amplitude at the qubit frequency $\fq$, immediately followed by a readout pulse applied at the resonator frequency $\fr$. Figure~\ref{Fig:Characterization}d displays the measurement of the relaxation time $T_1$, where we utilize a $\pi$-pulse (duration inferred from Fig.~\ref{Fig:Characterization}c) and vary the delay $t_\text{delay}$ between the end of each $\pi$-pulse and the onset of readout pulse. The population curve fitted by $\exp\left(-t/T_1\right)$ yields $T_1 = 15~\mu$s, which is long compared with most semiconductor charge qubits~\cite{Chatterjee2021}. We also verify the linear dependence of Rabi frequency $\Omega_{\mathrm{R}}$ on the pulse amplitude $V_\text{pulse}$ normalized by the maximally used amplitude $\bar{V}_\text{pulse}$ (see Fig.~\ref{Fig:Characterization}e).

Figure~\ref{Fig:Characterization}f shows the measurements of the original (Ramsey fringe) coherence time $T_2^*$ and extended (Hahn echo) coherence time $T_{2\text{E}}$~\cite{Krantz2019}. (See Methods for more details.) The Ramsey measurement consists of two $\pi/2$ pulses separated by a varied delay time $\Delta t$. The population curve is found to be best fitted by a Gaussian decay function $\exp[-(t/T_2^*)^2]$ with $T_2^*=50$~ns, which is consistent with the linewidth from the two-tone spectroscopy measurement and indicative of the probable dephasing from $1/f$ frequency noise~\cite{Ithier2005}. The Ramsey measurement is known to be sensitive to low-frequency electromagnetic fluctuations in the circuit. The Hahn echo measurement inserts an additional $\pi$ pulse in the middle point between two $\pi/2$ pulses. It mitigates low-frequency noise and transfers the decay function from Gaussian to exponential. Our Hahn echo population curve is best fitted with $\exp[-(t/T_{2\text{E}})^{1.5}]$~\cite{Chen2018}, which yields an extended (echo) coherence time $T_{2\text{E}} = 220$~ns. These results suggest that the qubit coherence may be primarily limited by low-frequency charge noise.

\noindent\textbf{Discussion and outlook}

The long $T_1$ manifests that solid Ne can indeed serve as an ultraclean substrate for single-electron qubits. We expect future trap geometries and better filtering for DC electrodes to give even longer $T_1$. The still short $T_2\ll T_1$ at this initial stage of development may originate from two sources of residual noise. First, remnant electrons along the resonator in the long channel are not entirely fixed; remaining motion can cause background charge noise to the trapped electron-photon interacting system. Second, Ne atoms on an imperfect (presumably rough and porous) surface can  be highly movable and induce a time-varying trapping potential to the electron. Improvement to the device design and Ne growth process~\cite{Sheludiakov2019}, and operation at the charge noise ``sweet spot" are expected to mitigate these decoherence issues. It has been theoretically calculated that the in-plane motional coherence of an electron on solid Ne surface can be several milliseconds~\cite{zavyalov2005electron}. Ultimately, utilizing the spin states through engineered spin-orbital coupling~\cite{Schuster2010,Mi2018} can yield ultralong qubit coherence in excess of 1~s~\cite{Lyon2006,Schuster2010,Sheludiakov2019,jin2020quantum}.

The strong interaction between the electron motional states and microwave photons will allow two or more electrons to entangle with each other through exchanging (virtual) photons in the resonator. To scale the system up, we can adopt the quantum charge-coupled device (QCCD) technique, originally developed trapped ion system~\cite{kielpinski2002,Bradbury2011,Pino2021}, to shuffle electrons into and out of different functional zones on a chip to achieve multi-electron gating, entanglement, and readout. This will significantly expand the scalability.

The eNe qubit platform incorporates compelling advantages from several leading qubit platforms; analogous to electromagnetically trapped ions, the electron qubits here are identically generated by a simple source and can have long spin coherence times; as with semiconductor quantum dots, electronic gate control can be applied at high speed; finally, strong coupling with the circuit QED architecture enables dispersive readout, transduction to microwave photons, and two-qubit gates via microwave resonator mediated interactions. Given these merits, we anticipate the eNe qubit platform to rapidly evolve into a superior quantum computing hardware. Furthermore, it can be coherently linked with other quantum information systems, {\eg, Josephson junctions and color centers through microwaves, to collectively advance quantum sensing, transduction, networks, and other important areas in quantum science, as well as fundamental physics.


%

\  \\
\noindent\textbf{Methods}

\noindent\textbf{Cryostat setup}

Our experiment is performed on a BlueFors LD400 dilution refrigerator system with a base temperature about 7~mK. Extended Data Fig.~1 shows our cryostat and measurement setup for single-electron qubits on solid neon (Ne) in a circuit quantum electrodynamics (QED) architecture. In this section, we focus on the cryostat setup inside the fridge. The measurement setup outside the fridge is explained in later sections when it is referred to.


All the input RF coaxes are made of silver-plated beryllium copper (SBeCu) from the room temperature (RT) plate (\ie, the 300~K plate) all the way to the mixing chamber (MXC) plate (\ie, the 10~mK plate). The output coaxes are made of SBeCu from the RT plate to the 3~K plate but superconducting niobium titanium (NbTi) from the 3~K plate to the MXC plate. Attenuators are installed along every RF line at every temperature stage to thermalize the cables and reduce the noise from RT. For the input lines, there is a 20~dB attenuation at 3~K, 10~dB at 1~K, 10~dB at 100~mK, and 20~dB at 10~mK. For the output lines, only 0~dB attenuation is used at every temperature stage for thermal anchoring. A 4--12~GHz cryogenic dual-junction isolator (Low-Noise Factory LNF-ISISC4\_12A) with 30~dB isolation is installed at the output of the sample cell to block the thermal noise from higher temperature. The isolator is followed by a 4--12~GHz cryogenic circulator (LNF-CIC4\_12A) with 50~$\Omega$ termination for another 20~dB isolation. A 4--8~GHz high-electron-mobility transistor (HEMT) amplifier (LNF-LNC4$\_$8C) with 39~dB gain and 2~K noise temperature is installed on the 3~K plate. Immediately outside the fridge, a 4--23~GHz RT low-noise amplifier (LNF-LNR4\_23A) with 27~dB gain and 65~K noise temperature (measured at 6~GHz) serves as the first-stage RT amplifier.


There are two types of DC ($\lesssim1$~kHz low-frequency) lines installed in our fridge: 1. thermocoaxes with inner core and outer shield made of stainless steel and dielectric filling made of magnesium oxide (MgO) powder; 2. twisted-pair DC wires consisting of phosphor bronze (PhBr) between the RT and 3~K plates and superconducting NbTi wires between the 3~K and MXC plates.

The DC voltages that control the electron trap and qubit detuning are delivered through thermocoaxes. They each provide $>35$~dB attenuation above 100~MHz. At the MXC plate, behind each thermocoax there is a two-stage RC low-pass filter with a cutoff frequency around 400~Hz cascaded with a LC low-pass filter (Mini-Circuits RLP-30+) with a cutoff frequency around 30~MHz. The voltage and current to generate electrons from the tungsten filaments are delivered through the twisted-pair wires.


A stainless steel fill line, consisting of two sections, is built in the fridge to fill Ne into a sample cell. The first section has 2~m length, 0.6~mm inner diameter, running from the RT plate down to beneath the 3~K plate, heat sunk at the 50~K plate and the 3~K plate. The second section has another 2~m length, 0.24~mm inner diameter, running from the end of the first section down to beneath the MXC plate, heat sunk on the still plate (\ie, the 1~K plate), the cold plate (\ie, the 100~mK plate), and the MXC plate. Then the fill line is converted into a brass flange, which interfaces by indium seal with another brass flange from the sample cell.


The sample cell consists of a copper lid and a copper pedestal, as shown in Extended Data Fig.~2. The lid contains 14 hermetic SMP feedthroughs (Corning 0119-783-1) for DC and RF signals, 2 SMP feedthroughs for the electron source, and a stainless steel tube (1/16" outer diameter and 0.021" inner diameter) for Ne filling. A custom designed printed circuit board (PCB) along with a 2$\times$7~mm sample chip is mounted on the pedestal inside the cell. The SMP connectors on the PCB are connected to the hermetic SMP connectors on the lid through SMP bullets (Rosenberger 19K106-K00L5). The lid and pedestal are sealed together by indium wires.


Two tungsten filaments are taken from standard 1.5~V miniature bulbs and mounted in parallel above the sample chip inside the sample cell (see Extended Data Fig.~2). The wire and coil diameter of the filament are 4~$\mu$m and 25~$\mu$m, respectively. There are about 30 coils for each filament. The resistance of each filament at RT is $15~\Omega$.

\noindent\textbf{Preparation experiments}


Extended Data Fig.~3 gives the phase diagram of neon. We use research-grade (99.999\% purity) Ne gas from Airgas. We fill Ne into the cell by first warming up the temperature of the 3~K plate to $25.6 - 26.4$~K and of the cell at the MXC plate to 25.5~K. The back-end pressure at the Ne tank is 10~psi above atmospheric pressure. Neon gas flows through a liquid nitrogen (LN$_2$) cold trap to remove any potential impurity that may clog the fill line. Then it reaches a volume control unit consisting of two solenoid valves (IMI Norgren U142010 24VDC), a pressure transducer (Swagelok PTI-S-AG60-22AQ) and a 10~cc cylinder (Swagelok SS-4CS-TW-10). Neon turns into liquid at the 3~K plate and drips into the cell along the second section of fill line. We estimate the amount of filled Ne by counting the number of puffs by repeatedly opening and closing the valves. Each puff corresponds to 10~cc Ne gas at room temperature.

The effective permittivity of the stripline resonator changes with the amount of Ne filled inside the channel. After filling, we gradually reduce the heating power to let the cell slowly cool across the triple point and then continue cooling down to 10~mK. By measuring the resonance frequency shift and comparing it with our numerical simulation, we can estimate the Ne thickness. In a typical experiment, we only send in about 40~puffs, which conformally coat about 5--10~nm solid Ne on the device surface and induce only 0.3--0.6~MHz frequency shift.

It is worthwhile to mention that even though solid Ne cryogenic substrate can exist at $\sim20$~K, operation of a circuit QED architecture still requires much lower temperatures to attain superconductivity and high-Q resonances. While we imagine that the device might operate at few Kelvin temperatures, these temperatures would correspond to a thermal environment for the $\sim6$~GHz frequencies of the qubit and resonator, making operation more challenging.


A vector network analyzer (VNA) (Keysight E5071C) is used to carry out the microwave transmission measurement through the superconducting stripline resonator. Port-1 and 2 of the VNA are connected respectively to the input and output RF lines of the device inside the fridge. This constitutes a standard $S_{21}$ measurement. We have done a separate reference measurement without the device and found that the total attenuation from Port-1 of VNA to the sample input to be about 70~dB, including the RT part of the cable loss.

The experimental transmission amplitude $(A/A_0)^2$ of the resonator is fitted with the Lorentzian function,
\begin{equation}
	(A/A_0)^2 = \frac{(\kappa/2\pi)^2}{4(f-\fr)^2+(\kappa/2\pi)^2},
\end{equation}
where $\fr$ is the resonator frequency and $\kappa/2\pi$ is the linewidth. The transmission phase $\phi$ is
\begin{equation}
	\phi = \arctan\frac{4\pi(f-\fr)}{\kappa}.
\end{equation}
The measured resonator frequency $\fr$ without Ne is $6.4266$~GHz. And the measured resonator linewidth $\kappa$ is $2\pi \times0.4$~MHz, giving a quality factor $Q\approx1.6\times10^4$. Fully filling the channel with Ne shifts the resonance to $6.2795$~GHz whereas 5--10~nm coating of the device surface only slightly shifts the resonance towards 6.426~GHz. With and without Ne, no significant change in $\kappa$ or $Q$ is observed.

The average photon occupancy $\bar{n}$ in the resonator is critical to our measurements. According to the input-output theory~\cite{walls2007quantum}, it can be estimated by
\begin{equation}
	\bar{n} = \frac{\kappa_\text{in}}{h f_\text{r} (\kappa_\text{in}+\kappa_\text{out}+\kappa_\text{i})^2}P_{\text{in}},
\end{equation}
where $P_{\text{in}}$ is the input power, $\kappa_\text{in}$ and $\kappa_\text{out}$ are the input and output coupling rates, and $\kappa_\text{i}$ is the intrinsic losses. For a typical overcoupled symmetric two-port superconducting stripline resonator, $\kappa_\text{in}\approx\kappa_\text{out}\gg\kappa_\text{i}$. Thus the overall linewidth $\kappa\approx\kappa_\text{in}+\kappa_\text{out}\approx 2\kappa_\text{in}$, and the average photon number can be simplified to
\begin{equation}
	\bar{n} = \frac{P_{\text{in}}}{2hf_\text{r} \kappa}. \label{Eqn:power}
\end{equation}
In our case, with the measured resonator frequency $\fr\approx 6.426$~GHz, the linewidth $\kappa/2\pi=0.4$~MHz, and the desired occupancy $\bar{n}\approx 1$, we can find the required input power $P_{\text{in}} \approx -135$~dBm.


Electrons are generated by applying voltage and current to the two filaments in parallel using a low-frequency function generator (Agilent 33220A). The driving pattern is a pulse train of 0.1~ms width, 4~V height, 1000~Hz frequency, for 1~s duration time.

A digital-to-analog converter (DAC) (NI-6363) provides all the DC voltages, including the resonator voltage $\Vr$, trap voltage $\Vt$, resonator-guard voltage $\Vrg$, and trap-guard voltage $\Vtg$, via thermocoaxes. When generating electrons through tungsten filaments, we fix $(\Vr, \Vt, \Vrg, \Vtg) = (1,0,0,0)$~V and monitor the evolution of the transmission spectrum with the VNA, as shown in Extended Data  Fig.~4. The positive $\Vr$ helps to attract electrons into the channel. After the initial electron injection, the spectrum undergoes a sudden change with a large resonance frequency shift. After a few seconds, the resonance peak restores back close to the starting frequency. However, too many electrons may have been deposited onto the trap and resonator in the channel. To remove most of the redundant electrons, we reverse $\Vr$ to a large negative voltage to repel electrons away. This eventually brings the resonance frequency back very close to the bare resonator frequency with very few electrons remained on the trap and resonator. At this point, we switch $\Vr$ back to $+1$~V to keep unwanted electrons in the channel far off resonance with the resonator and thereby minimize the charge noise from the channel.


We use the procedure given in our previous work to load individual electrons from the resonator (reservoir) onto the trap~\cite{Koolstra2019}. This procedure involves a complex sequence of DC voltage tuning. During this procedure we keep watching the change of microwave transmission from the VNA at the bare resonator frequency $\fr$. Since the number and position of electrons in the channel are not completely deterministic, we cannot yet guarantee every loading process to capture a single electron. Sometimes, we have to repeat the electron generation process. However, our overall successful rate is very high.

After each loading procedure, we fix the resonator voltage $\Vr=1$~V and trap-guard voltage $\Vtg=0$~V and fine scan the resonator-guard voltage $\Vrg$ and trap voltage $\Vt$ in a wide range spanning several hundreds of mV. If a single electron is indeed present, we can observe one or two sharp absorption lines. Those lines correspond to the case when the qubit frequency $\fq$ matches the resonator frequency $\fr$ and induce transmission absorption. Extended Data Fig.~5 displays the observed absorption lines corresponding to the electron qubit presented by Fig.~3 in the main paper.

\noindent\textbf{Frequency-domain measurements}

In the vacuum Rabi splitting measurement, we keep the VNA setup unchanged from the transmission measurement above.
The probe frequency $\fp$ of the VNA is kept sweeping around the bare resonator frequency in a narrow range $\fr\pm10$~MHz, while the resonator-guard voltage $\Vrg$ is tuned by $\pm 1$~mV from the on-resonance value. All other voltages are fixed. In the plot of normalized transmission amplitude $(A/A_0)^2$ versus $\fp$ and $\Vrg$, the avoided crossing is sharpest when the qubit and resonator are on-resonance coupled. The splitting between the two transmission peaks in this on-resonance condition yields the coupling constant $2g/2\pi$. In this measurement, to avoid power broadening, the probe power is kept low according to Eq.~(\ref{Eqn:power}) so that the average photon occupancy $\bar{n}$ inside the cavity is about 1.

The vacuum Rabi splittings corresponding to the electron qubit presented in Fig.~3 of the main paper is shown in Extended Data Fig.~6. According to the input-output theory~\cite{walls2007quantum}, the transmission spectrum of the coupled system reads
\begin{equation}
	S_{21} = \frac{\kappa/2}{i2\pi(\fr-\fp)+\kappa/2+ig^2/( 2\pi\fq-2\pi\fp-i\gamma)},
\end{equation}
under the assumption that the resonator intrinsic decay rate $\kappa_\text{i}$ is much less than the input and output coupling rates $\kappa_\text{in}=\kappa_\text{out}=\kappa/2$. With this formula, the fitted coupling strength is $g=2\pi \times$4.5~MHz and the electron linewidth is $\gamma = 2\pi \times3.4$~MHz. The latter is slightly larger than that obtained through the two-tone qubit spectroscopy.

In the two-tone qubit spectroscopy measurement, the VNA provides the first tone, \ie, the probe tone, and a signal generator (Anritsu MG3692C) provides the second tone, \ie, the pump tone. The qubit spectrum is tuned by the resonator-guard voltage $\Vrg$. The VNA setup is the same as before. Port-1 and 2 of the VNA are connected to the input and output lines of the device, but the probe frequency is fixed only at the bare resonator frequency $\fr=6.426$~GHz and is not swept. The signal generator generates a continuous wave at the pump frequency $\fs$ that is swept over a broad range from $\fr-1.2$~GHz to $\fr+0.2$~GHz. This pump tone is combined with the signal from Port-1 of VNA and sent into the input line of the device inside the fridge. The combination is made through a directional coupler. The pump tone goes into the coupled port of the directional coupler with $-10$~dB signal reduction. For each given $\Vrg$, $\fs$ is swept to produce the qubit spectrum for that particular $\Vrg$. Then $\Vrg$ is scanned over a range of several hundreds of mV. The transmission phase at $\fr$ is recorded as a function of both $\Vrg$ and $\fs$, and generates a complete qubit spectrum.

When the pump frequency $\fs$ is far detuned from the resonator frequency $\fr$, the resonator suppresses the pump amplitude and increases the required power. In our experiments, we apply $-10$~dBm pump power from the signal generator. Taking account of the $70$~dB attenuation on the input line and 10~dB coupling of the directional coupler, the input pump power at the sample is $-90$~dBm. The measured qubit linewidth is dependent on the pump power, known as power broadening. Practically, we vary the power to attain the narrowest linewidth while maintaining a reasonable signal-to-noise ratio.

In order to see higher order transitions and gain some insight about the anharmonicity of this qubit, we pump the system even harder. We are able to see many other transitions in the range of $\pm100$~MHz detuning, which is the range of our main interest. Extended Data Fig.~7 shows the two-tone spectroscopy with high power qubit pumping. At $-100$~MHz detuning, the anharmonicity between the two lowest transition frequencies is $\alpha/2\pi \equiv f_{|1\rangle\rightarrow|2\rangle} - f_{|0\rangle\rightarrow|1\rangle} \approx 40$~MHz, where $|0\rangle$, $|1\rangle$, $|2\rangle$ are the ground, first-excited, and second-excited states, respectively. This number can provide an estimate of the frequency shift in dispersive readout.

\noindent\textbf{Time-domain measurements}

In our time-domain measurements, the qubit frequency is tuned to $\fq=6.326$~GHz by the DC electrodes. Compared with the bare resonator frequency at $\fr = 6.426$~GHz, the qubit-to-resonator detuning is $\Delta/2\pi = \fq - \fr = -100$~MHz.

The following procedure is used to generate qubit gate pulses. The signal generator (Anritsu MG3692C) generates a continuous sine wave at $6.206$~GHz, which is $120$~MHz below $\fq$. It is sent into the LO port of a IQ-mixer (MarkiMMIQ 0520H). An arbitrary waveform generator (AWG) (Tektronics AWG5204) generates, at its Channel 1 and 2, two rectangle-enveloped sine-wave pulse sequences, with the same amplitude, pulse length, and 120~MHz central frequency, but different phases offset by $\pi/2$ from each other. They are respectively sent into the two IF ports, \ie, the I and Q ports of the IQ-mixer. The output from the RF port of the IQ-mixer contains only the upper sideband pulse sequence with the carrier frequency at the qubit frequency $\fq = 6.326$~GHz. The lower sideband and the LO sine wave are both suppressed.

The following procedure is used to generate qubit readout pulses. Another signal generator (Lab Brick LMS-183DX) generates a continuous sine wave at the bare resonator frequency $\fr = 6.426$~GHz. It is sent into the LO port of a mixer (MiniCircuit ZMX-8GH). The AWG generates, from its Channel 3 that is synchronized and delayed from Channel 1 and 2, a rectangle-enveloped DC pulse sequence with appropriate amplitude and typically $2~\mu$s pulse length. It is sent into the IF port of the mixer. The output from the RF port of this mixer is a rectangle-enveloped pulse sequence with the carrier frequency at the resonator frequency $\fr=6.426$~GHz. Instead of being sent straightly into the fridge, the readout pulse sequence is first sent into the input port of a directional coupler and splits off only a $-20$~dB portion from the coupler port to go into the fridge. The readout pulse from the output port of the directional coupler serves as a reference signal in the dispersive phase measurement setup.

The qubit gate pulse sequence and the readout pulse sequence are combined through a combiner and sent into the input line of the circuit QED device inside the fridge. Although they travel along the same line, they do not overlap in real time. The transmitted readout pulse sequence, containing the phase information from the dispersive coupling with the qubit, is then routed to the RF port of a second mixer (MiniCircuit ZMX-8GH) for heterodyne detection. A third signal generator (Lab Brick LMS-183DX) provides a continuous sine wave at $\fr + 50$~MHz to the LO port of this mixer. The signal from the IF port is further amplified and sent into Channel 2 of a digitizer (Alazar ATS9870). The reference readout pulse from the output port of the directional coupler is routed into the RF port of a third mixer (MiniCircuit ZMX-8GH). The same third signal generator provides the same continuous sine wave to the LO port of this mixer through a splitter. The signal from the IF port of this mixer is sent into Channel 1 of the digitizer.

All the time-domain measurements, including Rabi oscillations, $T_1$ relaxation time and $T_2$ coherence time measurements follow exactly the same scheme as in superconducting qubit measurements~\cite{Krantz2019} and have been elaborated in the main paper. For all the time-domain measurements, statistical results are obtained with 5000 averages.

\noindent\textbf{Theoretical analysis}

Calculation on the electronic states along the $z$ direction follows the standard procedure in numerically solving a one-dimensional Schr\"odinger equation~\cite{jin2020quantum}. In the limit of barrier height $U\rightarrow\infty$ and short-range cutoff $b\rightarrow 0$, the problem can be mapped into that of the radial part of a hydrogen atom and yields analytical solution~\cite{Cole1969,Cole1971}. However, compared with an electron on liquid helium (eHe), an electron on solid neon (eNe) experiences a 30~\% lower repulsive barrier but 4 times stronger attractive potential, and is much more tightly bound to the surface at only 1--2~nm mean distance. Therefore,  analytical solutions can have large errors. The results presented in Fig.~1a of the main paper come from our finite-difference numerical calculation at an extremely fine space step of $0.1$~{\AA}. However, even with such a fine step, there are still several important points to be kept in mind: 1. The calculated eigen-energies are still sensitive to the accurate choice of $U$ and $b$ due to the divergent nature of the $-1/z$ like polarization potential as $z\rightarrow0$. There is always a competition between the positive part of the potential at $z\lesssim0^-$ and negative part at $z\gtrsim0^+$ which determines how much wavefunction ``spills" into the barrier. A tiny difference in ``spilling", even not so discernible by looking at wavefunctions, can cause a large difference in energy. 2. Solid neon has an fcc crystal structure with the lattice constant $a=4.464$~{\AA}, which is not only an order-of-magnitude larger than our choice of numerical space step but also larger than the standard choice of cutoff $b\approx \frac{1}{2}a=2.3~\text{\AA}$~\cite{Cole1969,Cole1971}. It is not yet clear at what length scale the continuous model used here breaks down and how critically the exact atomic arrangement and surface profile influence the qualitative picture. These questions can in principle be addressed by advanced computation methods such as quantum Monte Carlo (QMC) and eventually answered only by experiments.

Here we discuss more about the electronic states in the $xy$ plane that are directly related to our experiment. The observed single-electron qubit spectrum in Fig.~3a of the main paper exhibits a quadratic (parabolic or hyperbolic) curve that is symmetric with respect to a ``sweet spot" voltage, $\Vss=339$~mV, in the detuning range of the resonator-guard voltage $\Vrg$. This feature qualitatively resembles that of a semiconductor double-quantum-dot (DQD) spectrum~\cite{Mi2017}. The obtained Extended Data Fig.~5 and Fig.~6 for electron-photon coupling and vacuum Rabi splitting also look very similar to those of a DQD qubit~\cite{Mi2017}. While our electron trapping potential was not intended to be a double quantum well, it is actually not a surprise to observe a DQD spectrum, as elaborated below.

Our device was designed to be symmetric with respect to the $y=0$ plane. In the ideal situation, if an electron happens to be trapped at the center of the trap and solid neon is grown perfectly flat there, the in-plane trapping potential, after a Taylor expansion in $y$ should contain only even-order terms, $ V(y) \approx V_0 + \frac{1}{2} k_2 y^2 + \frac{1}{4!} k_4 y^4 + \cdots$.
The coefficients like $V_0$, $k_2$ and $k_4$ can be complicated functions of all the DC voltages. The Taylor expansion along $x$ may contain odd terms like $x$ and $x^3$ terms and cross terms to $y$ like $x^2y^2$. However, since we have arranged to trap electron more tightly along $x$ on its ground state, we can completely ignore $x$ here. Specific to our device configuration shown in Fig.~1 of the main text and Extended Data Fig.~1, if the two arms of the resonator-guard electrodes applied with the voltage $\Vrg$ are indeed perfectly symmetric, then when we slightly tune $\Vrg$ and keep everything else fixed, the leading consequences are the change of overall potential minimum $V_0$ and the linear coefficient along $x$, but not $k_2$ or $k_4$, and hence should have minimal influence to the transition spectrum associated with the electron's $y$-motion.

However, if the device has some small asymmetry due to fabrication imperfection, defected solid Ne surface, or background electrons in different experiments, then the Taylor expansion of trapping potential along $y$ must include the linear and cubic terms. At the perturbative level, we can keep just the linear term, drop the spectrally irrelevant $V_0$ term, and rewrite the potential as
\begin{equation}
	V(y) \approx \frac{1}{2}k_2\left[ \beta y + y^2 + \zeta y^4 \right] .
\end{equation}
The parameter $\beta$, with the dimension of length, can be called the asymmetry parameter. The parameter $\zeta$, with the dimension of length$^{-2}$, can be called the anharmonicity parameter. If $\beta\rightarrow0$, the systems goes back to a symmetric anharmonic oscillator up to the quartic term. And if $\zeta\rightarrow0$ too, the system returns to a simple harmonic oscillator with $k_2$ as the usual spring constant. All the parameters $k_2$, $\beta$ and $\zeta$ are in principle functions of the tuning voltage $\Vrg$ here. But for small tuning, $\beta$ is the leading variable whereas $k_2$ and $\zeta$ are approximately constant. At the perturbative level, we can relate $\beta$ and $\Vrg$ by $\beta = (\Vrg-\Vss)/\eta$, where $\eta$ is a positive constant and has the physical meaning of a characteristic electric field, and $\Vss=339$~mV is the aforementioned ``sweet spot" voltage. If $\Vrg=\Vss$, then $\beta=0$, $V(y)$ is symmetric. If $\Vrg>\Vss$, then $\beta>0$, $V(y)$ is slightly higher on the right and so electron is slightly shifted to the left. If $\Vrg<\Vss$, then $\beta<0$, $V(y)$ is slightly higher on the left and so the electron is slightly shifted to the right.  Therefore, the minimal model that can capture most of the experimental features reads
\begin{equation}
	V(y) = \frac{1}{2}k_2\left[ \frac{\Vrg-\Vss}{\eta} y + y^2 + \zeta y^4 \right] ,
\end{equation}
where $k_2$, $\eta$ and $\zeta$ are all constants to be determined from experiments. With the choice of $k_2=5.536$~meV~$\mu$m$^{-2}$, $\eta = 0.8271$~V~$\mu$m$^{-1}$ , $\zeta=15.4$~$\mu$m$^{-2}$, the obtained qubit spectrum and anharmonicity can largely reproduce the experimental results.

Extended Data Fig.~8 presents all the calculated qubit properties based on the minimal model above. The trapping potential is flattened at the bottom by the anharmonic quartic term and symmetrically leans to the left and right by tuning $\Vrg$ with respect to $\Vss$. Electron wavefunctions extend about 500~nm in space and are left and right shifted with the potential changes. In particular, the calculated $|0\rangle\rightarrow|1\rangle$ transition spectrum in Extended Data Fig.~8e is nearly identical to Fig.~3a in the main paper from the two-tone qubit spectroscopy measurement. The magnified spectrum in the $\pm100$~MHz detuning range gives the $|1\rangle\rightarrow|2\rangle$ transition with a positive anharmonicity $\alpha=40$~MHz at $-100$~MHz detuning, in agreement with Extended Data Fig.~7.

It is known that for a symmetric potential, if only the lowest three energy levels need to be considered, the dispersive shift can be theoretically estimated by $\chi\approx g^2\alpha/\Delta(\Delta+\alpha)$, as is used most often in superconducting transmon qubits. However, for a generally asymmetric potential, the originally forbidden transitions by selection rules are activated and higher-lying levels often need to be included in order to get an accurate agreement with experiments.

Taking our case as an example, if we assume the asymmetry is not too drastic and the lowest three levels still dominate the main behaviors, the dispersive shift can be approximated by~\cite{DavidIsaacSchuster2007}
\begin{equation}
	\chi=\chi_{01}-\frac{\chi_{12}}{2} + \frac{\chi_{02}}{2},
\end{equation}
with contributions from each $i\rightarrow j$ transition (including the previously forbidden $0\rightarrow 2$ transition),
\begin{equation}
	\chi_{ij}  = \frac{g_{ij}^2}{\omega_{ij}-\omega_{\text{r}}} = \frac{g_{ij}^2}{\Delta_{ij}}.
\end{equation}
Here $g_{ij}$ is the generalized coupling strength associated with the $i\rightarrow j$ transition, $\omega_{ij}=\omega_j - \omega_i$ is the transition frequency, $\omega_{\text{r}}$ is the resonator frequency, and $\Delta_{ij} = \omega_{ij} - \omega_{\text{r}}$ is the generalized detuning. $g_{ij}$ is proportional to the electric dipole strength $d_{ij}$ at the transition frequency and normalized single-photon (zero-point) electric field strength $\mathcal{E}_{\text{r}}$ of the resonator mode.

Further analysis requires calculating the dipole strengths from the approximate electron wavefunctions obtained from the minimal model above and estimating the single-photon (zero-point) field strength from the vacuum Rabi splitting measurement. However, we do not expect such a simple treatment can lead to a highly meaningful comparison between theory and experiment. We intend to carry out more systematic studies of the single-electron qubit spectroscopy, both experimentally and theoretically, in our future works.

\ \\
\noindent\textbf{Data availability}

The data that support the findings of this study are available from the corresponding authors upon request.

\ \\
\noindent\textbf{Code availability}

The computer codes that are used in this study are available from the corresponding authors upon request.

\ \\
\noindent\textbf{Acknowledgements}

This work was performed at the Center for Nanoscale Materials, a U.S. Department of Energy Office of Science User Facility, and supported by the U.S. Department of Energy, Office of Science, under Contract No. DE-AC02-06CH11357. D. J. acknowledges additional support from the Julian Schwinger Foundation (JSF) for Physics Research for hardware component upgrade. X. L. acknowledges additional support from Argonne National Laboratory Directed Research and Development (LDRD) Program for device characterization effort. This work was partially supported by the University of Chicago Materials Research Science and Engineering Center, which is funded by the National Science Foundation under award number DMR-2011854. This work made use of the Pritzker Nanofabrication Facility of the Institute for Molecular Engineering at the University of Chicago, which receives support from SHyNE, a node of the National Science Foundations National Nanotechnology Coordinated Infrastructure (NSF NNCI-1542205). D. I. S. and B. D. acknowledge support from NSF Grant No. DMR-1906003. K. W. M acknowledges support from NSF Grant No. PHY-1752844 (CAREER) and use of facilities at the Institute of Materials Science and Engineering at Washington University. W. G. acknowledges support from NSF Grant No. DMR-2100790 and the National High Magnetic Field Laboratory, which is funded through the NSF Cooperative Agreement No. DMR-1644779 and the State of Florida. G. Y. acknowledges support from the National Science Foundation under Cooperative Agreement PHY-2019786 (the NSF AI Institute for Artificial Intelligence and Fundamental Interactions, http://iaifi.org/). D. J. thanks Milton W. Cole, Mark I. Dykman, Stephen K. Gray, Paul Leiderer, Daniel Lopez, and Tijana Rajh for inspiring discussions. The authors thanks MIT Lincoln Laboratory and Intelligence Advanced Research Projects Activity (IARPA) for providing the traveling-wave parametric amplifier (TWPA) used in this project.

\ \\
\noindent\textbf{Author contributions}

X. Zhou and D. J. devised the experiment and wrote the manuscript. X. Zhou performed the experiment. G. K., G. Y., and D. I. S. designed the device. G. K. and G. Y. fabricated the device. X. Zhou, X. Zhang, X. H., and D. J. built the experimental setup. B. D. simulated the device. X. L. and R. D. characterized the device. W. G. advised the sample processing and theoretical modeling. K. W. M. and D. I. S. advised the measurement and revised the manuscript. D. J. conceived the idea and led the project. All authors contributed to the manuscript.

\ \\
\noindent\textbf{Competing interests}

Authors declare no competing interests.



\clearpage

\setcounter{figure}{0}
\makeatletter
\renewcommand{\figurename}{Extended Data Fig.}
\renewcommand{\fnum@figure}[1]{\textbf{\figurename~\thefigure~\textbar} }
\makeatother

\begin{figure*}
\centerline{\includegraphics[scale=1]{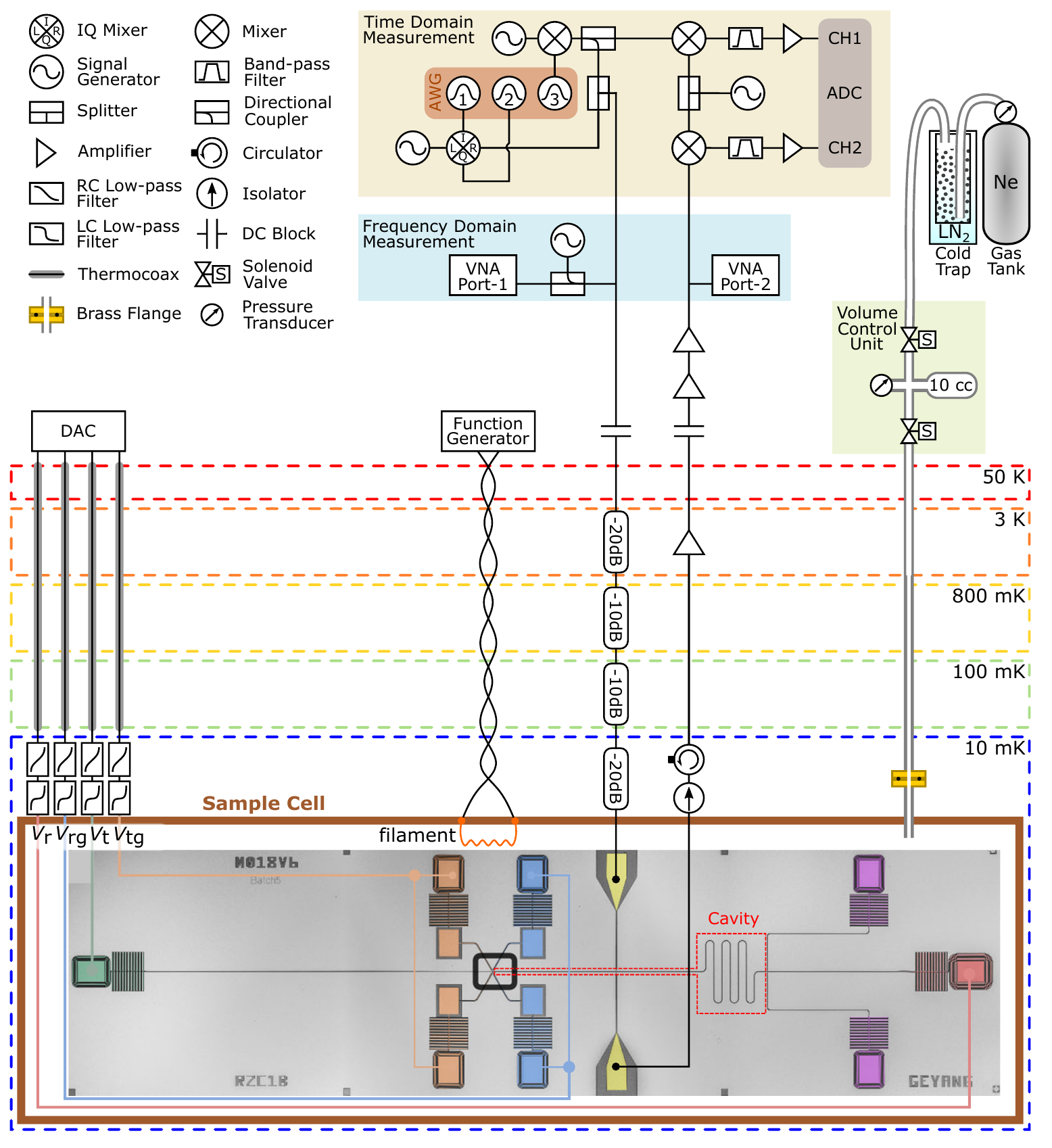}}
\caption{Cryostat and measurement setup for single-electron qubits on solid neon in a circuit quantum electrodynamics architecture. Details are explained in the text where they are referred to.}
\end{figure*}

\begin{figure*}
\centerline{\includegraphics[scale=1]{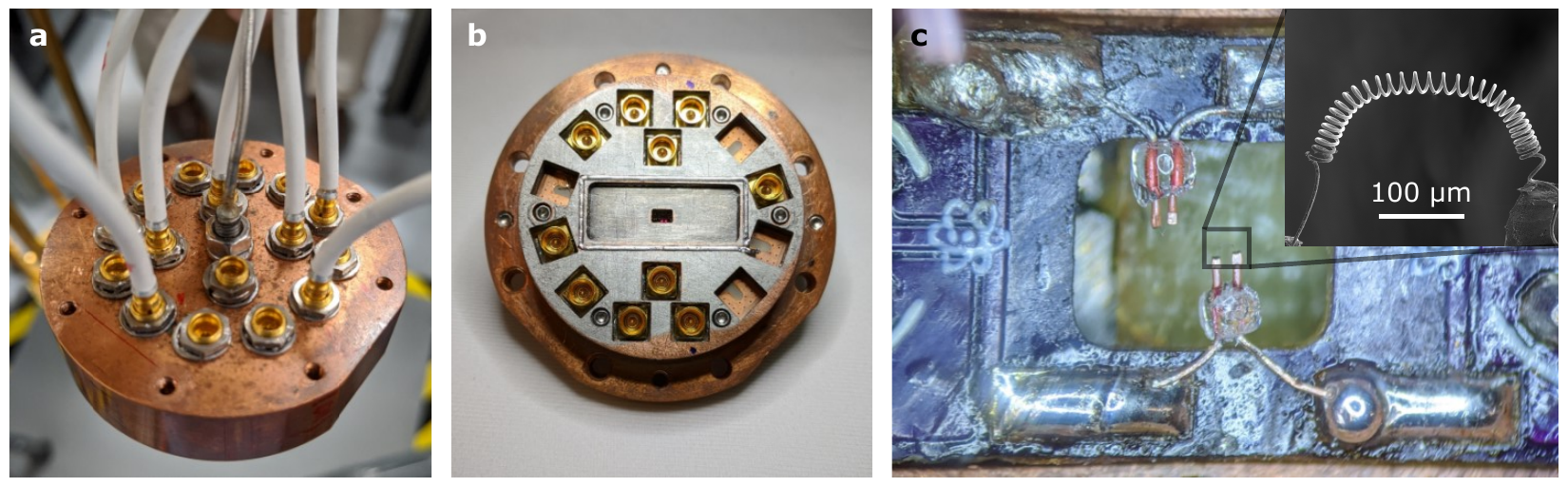}}
\caption{Photographs of sample cell and electron source. \textbf{a}, Lid part of the cell with all the coax connection. It contains 14 hermetic SMP feedthroughs for DC and RF signals, 2 SMP feedthroughs for electron source, and a stainless steel tube for neon filling. \textbf{b}, Pedestal part of the cell with a printed circuit board (PCB) mounted underneath a stack of copper sheets that suppress unwanted microwave modes. \textbf{c}, Two tungsten filaments, mounted in parallel on the back side of the lid in (\textbf{a}), as the electron source by thermionic emission. The inset shows a scanning electron microscopy (SEM) image of one tungsten filament.}
\end{figure*}

\begin{figure*}
\centerline{\includegraphics[scale=1]{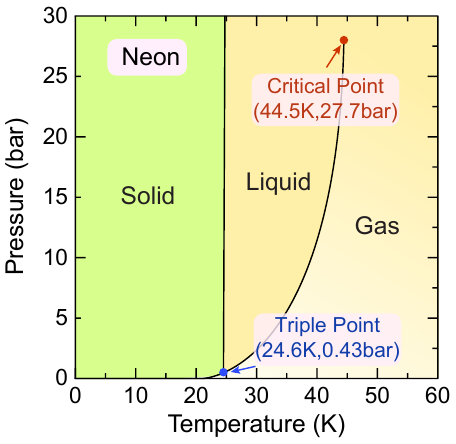}}
\caption{Phase diagram of neon. The solid-liquid-gas triple point is at $(24.56~\text{K},0.43~\text{bar})$ and the liquid-gas critical point is at $(44.49~\text{K},27.69~\text{bar})$.}
\end{figure*}

\begin{figure*}
\centerline{\includegraphics[scale=1]{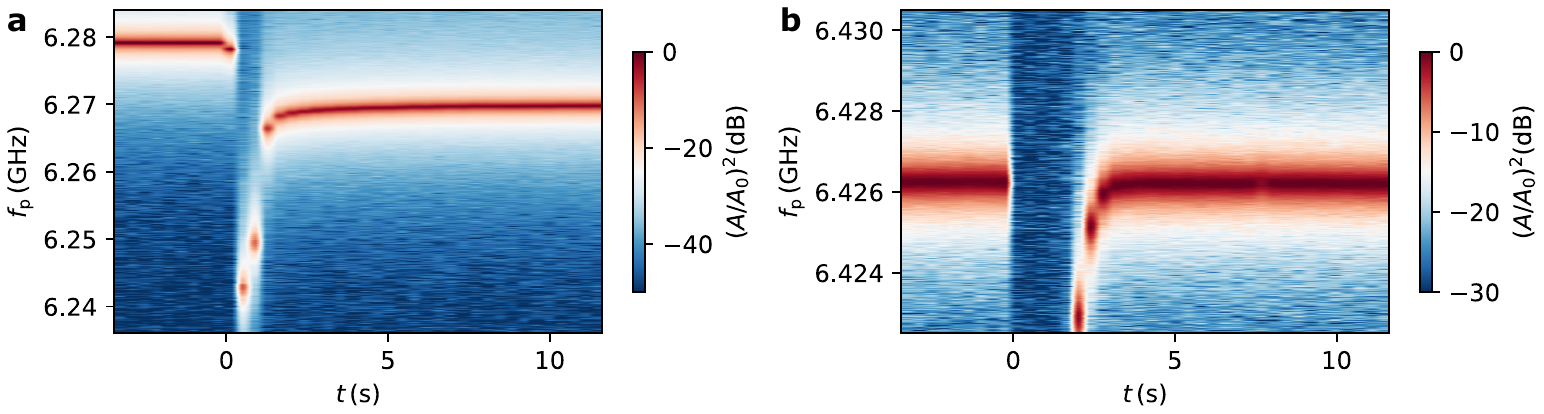}}
\caption{Observed time evolution of transmission amplitude $(A/A_0)^2$ during the electron generation and deposition processes. \textbf{a}, In the case of neon fully filling the channel. \textbf{b}, In the case of 5--10~nm neon conformally coating the device. At $t=0$, pulse train is sent to the tungsten filaments and electrons are generated and deposited onto the resonator. A sudden change in the spectrum can be seen. After about 3~s, the spectrum stabilizes and shows a frequency shift about 10~MHz for (\textbf{a}) and almost no shift for (\textbf{b}).}
\end{figure*}

\begin{figure*}
\centerline{\includegraphics[scale=1]{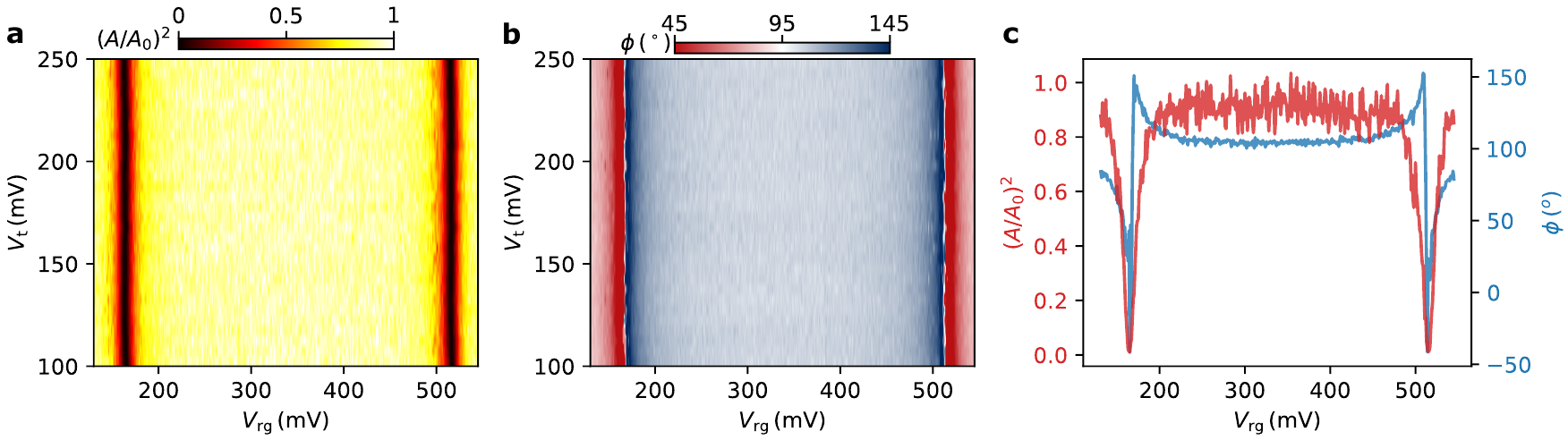}}
\caption{Coupling of a single electron and microwave photons. \textbf{a}, Normalized transmission amplitude $(A/A_0)^2$ probed at the bare resonator frequency $\fr$ as a function of the resonator-guard voltage $\Vrg$ and the trap voltage $\Vt$. \textbf{b}, Transmission phase $\phi$, corresponding to the amplitude in (\textbf{a}), as a function of  $\Vrg$ and $\Vt$. \textbf{c}, Line scanned normalized amplitude $(A/A_0)^2$ and phase $\phi$ as a function of $\Vrg$ at $\Vt=175$~mV. A dip in amplitude and 2$\pi$ phase jump occur when the qubit frequency matches the resonator frequency.}
\end{figure*}

\begin{figure*}
\centerline{\includegraphics[scale=1]{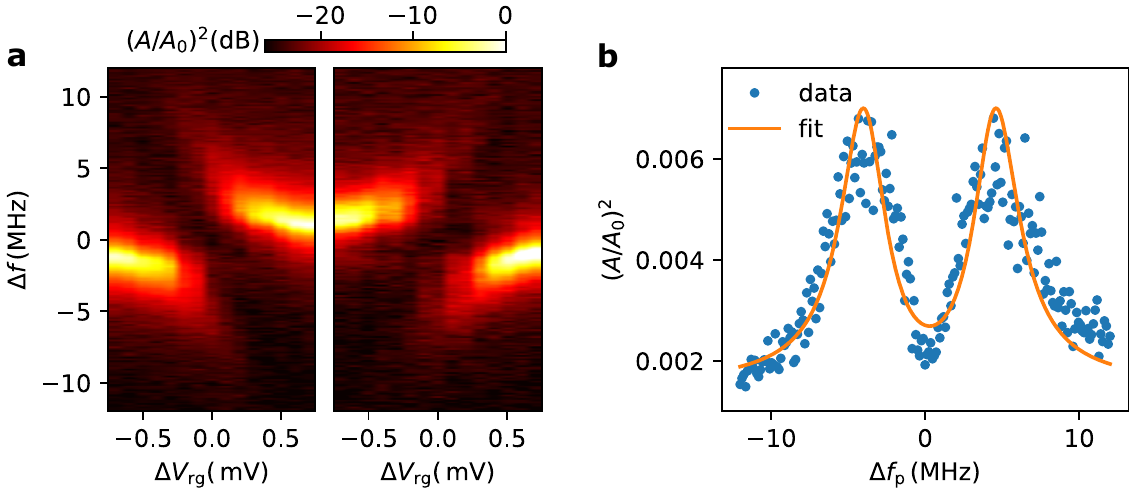}}
\caption{Vacuum Rabi splitting between a single electron and microwave photons. \textbf{a}, Normalized transmission amplitude $(A/A_0)^2$ as a function of probe frequency $\Delta f_\mathrm{p}=\fp-\fr$ and resonator-guard voltage $\Delta V_\mathrm{rg}$ (detuning from the resonance condition). \textbf{b}, Transmission amplitude $(A/A_0)^2$ versus a probe frequency when qubit and resonator is on resonance. The fitting curve with input-output theory gives a coupling strength  $g/2\pi$ about 4.5\,MHz and qubit decay rate $\gamma/2\pi$ about 3.4~MHz.}
\end{figure*}

\begin{figure*}
\centerline{\includegraphics[scale=1]{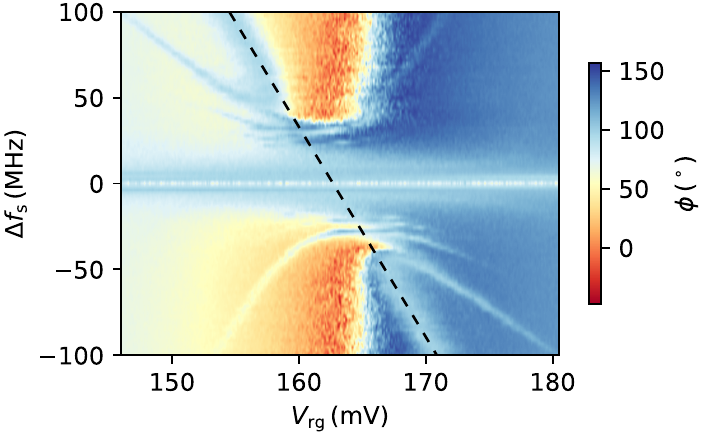}}
\caption{Two-tone qubit spectroscopy measurement with high pump power and pump frequency around the bare resonator frequency. Besides the $|0\rangle\rightarrow|1\rangle$ transition line, which is marked with black dashed line, there are other transition lines visible. The line immediately next to the main transition line is the $|1\rangle\rightarrow|2\rangle$ transition. At $\Delta/2\pi=\fq-\fr = -100$~MHz detuning, the anharmonicity $\alpha/2\pi \equiv f_{|1\rangle\rightarrow|2\rangle} - f_{|0\rangle\rightarrow|1\rangle} \approx 40$~MHz.}
\end{figure*}

\begin{figure*}
\centerline{\includegraphics[scale=1]{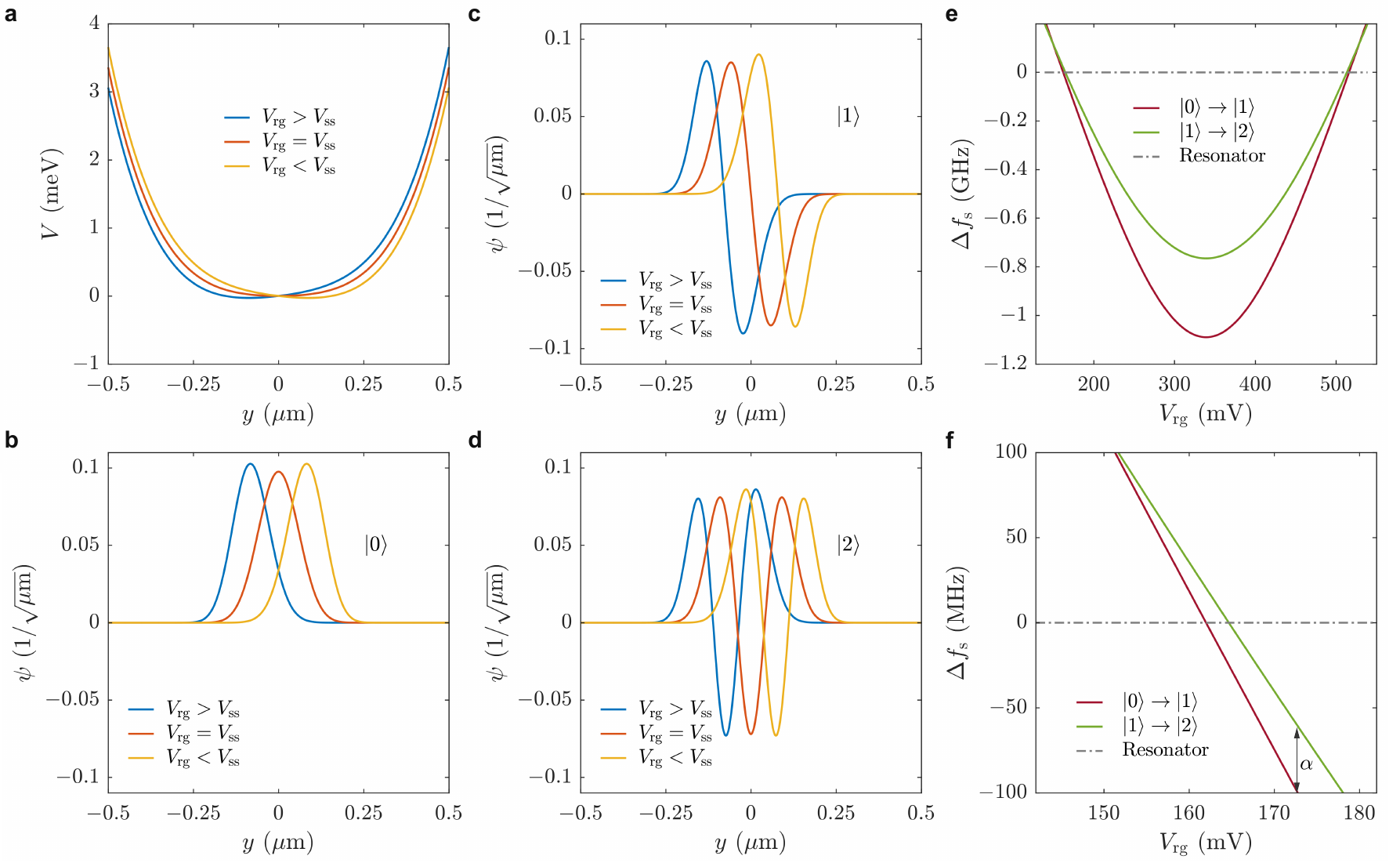}}
\caption{Calculated electron qubit properties based on a minimal model that encloses linear asymmetry and quartic anharmonicity. \textbf{a}, Trapping potential $V$ versus position $y$. The shape symmetrically leans to the left and right by tuning the resonator-guard voltage $\Vrg$ with respect to the ``sweet spot" voltage $\Vss=339$~mV. For $\Vrg>\Vss$, we take $\Vrg=$~516~mV and for $\Vrg>\Vss$, we take $\Vrg=$~162~mV, both of which are on-resonance conditions in experiment when the qubit frequency $\fq$ matches the resonator frequency $\fr$. \textbf{b}--\textbf{d}, Electron wavefunctions on the ground state $|0\rangle$, first excited state $|1\rangle$, and second excited state $|2\rangle$, respectively, for the three different $\Vrg$'s. They extend about 500~nm in space and are left and right shifted with the potential changes. \textbf{e}, Qubit spectrum under frequency scanning $\Delta f_\text{s} = f_\text{s}-\fr$ and voltage $\Vrg$ detuning, for $|0\rangle\rightarrow|1\rangle$ and $|1\rangle\rightarrow|2\rangle$ transitions . The first transition (in red) matches well with the experimental observation shown in Fig.~3a. \textbf{f}, Magnified qubit spectrum of $(\textbf{e})$ in the $\pm100$~MHz detuning range. The second transition has a positive anharmonicity $\alpha = 40$~MHz above the first transition at $-100$~MHz detuning. The overall spectral profile also matches the experiment observation shown in Extended Data Fig.~7, taking account of the practical spectrum deformation due to the overly strong pumping near resonance.}
\end{figure*}


\begin{thebibliography}{65}%
	\makeatletter
	\providecommand \@ifxundefined [1]{%
		\@ifx{#1\undefined}
	}%
	\providecommand \@ifnum [1]{%
		\ifnum #1\expandafter \@firstoftwo
		\else \expandafter \@secondoftwo
		\fi
	}%
	\providecommand \@ifx [1]{%
		\ifx #1\expandafter \@firstoftwo
		\else \expandafter \@secondoftwo
		\fi
	}%
	\providecommand \natexlab [1]{#1}%
	\providecommand \enquote  [1]{``#1''}%
	\providecommand \bibnamefont  [1]{#1}%
	\providecommand \bibfnamefont [1]{#1}%
	\providecommand \citenamefont [1]{#1}%
	\providecommand \href@noop [0]{\@secondoftwo}%
	\providecommand \href [0]{\begingroup \@sanitize@url \@href}%
	\providecommand \@href[1]{\@@startlink{#1}\@@href}%
	\providecommand \@@href[1]{\endgroup#1\@@endlink}%
	\providecommand \@sanitize@url [0]{\catcode `\\12\catcode `\$12\catcode
		`\&12\catcode `\#12\catcode `\^12\catcode `\_12\catcode `\%12\relax}%
	\providecommand \@@startlink[1]{}%
	\providecommand \@@endlink[0]{}%
	\providecommand \url  [0]{\begingroup\@sanitize@url \@url }%
	\providecommand \@url [1]{\endgroup\@href {#1}{\urlprefix }}%
	\providecommand \urlprefix  [0]{URL }%
	\providecommand \Eprint [0]{\href }%
	\providecommand \doibase [0]{https://doi.org/}%
	\providecommand \selectlanguage [0]{\@gobble}%
	\providecommand \bibinfo  [0]{\@secondoftwo}%
	\providecommand \bibfield  [0]{\@secondoftwo}%
	\providecommand \translation [1]{[#1]}%
	\providecommand \BibitemOpen [0]{}%
	\providecommand \bibitemStop [0]{}%
	\providecommand \bibitemNoStop [0]{.\EOS\space}%
	\providecommand \EOS [0]{\spacefactor3000\relax}%
	\providecommand \BibitemShut  [1]{\csname bibitem#1\endcsname}%
	\let\auto@bib@innerbib\@empty
	\bibitem [{\citenamefont {Ladd}\ \emph {et~al.}(2010)\citenamefont {Ladd},
		\citenamefont {Jelezko}, \citenamefont {Laflamme}, \citenamefont {Nakamura},
		\citenamefont {Monroe},\ and\ \citenamefont {O'Brien}}]{Ladd2010}%
	\BibitemOpen
	\bibfield  {author} {\bibinfo {author} {\bibfnamefont {T.~D.}\ \bibnamefont
			{Ladd}}, \bibinfo {author} {\bibfnamefont {F.}~\bibnamefont {Jelezko}},
		\bibinfo {author} {\bibfnamefont {R.}~\bibnamefont {Laflamme}}, \bibinfo
		{author} {\bibfnamefont {Y.}~\bibnamefont {Nakamura}}, \bibinfo {author}
		{\bibfnamefont {C.}~\bibnamefont {Monroe}},\ and\ \bibinfo {author}
		{\bibfnamefont {J.~L.}\ \bibnamefont {O'Brien}},\ }\bibfield  {title}
	{\enquote {\bibinfo {title} {{Quantum computers}},}\ }\href@noop {}
	{\bibfield  {journal} {\bibinfo  {journal} {Nature}\ }\textbf {\bibinfo
			{volume} {464}},\ \bibinfo {pages} {45--53} (\bibinfo {year}
		{2010})}\BibitemShut {NoStop}%
	\bibitem [{\citenamefont {Popkin}(2016)}]{Popkin2016}%
	\BibitemOpen
	\bibfield  {author} {\bibinfo {author} {\bibfnamefont {G.}~\bibnamefont
			{Popkin}},\ }\bibfield  {title} {\enquote {\bibinfo {title} {{Quest for
					Qubits}},}\ }\href@noop {} {\bibfield  {journal} {\bibinfo  {journal}
			{Science}\ }\textbf {\bibinfo {volume} {354}},\ \bibinfo {pages} {1090--1093}
		(\bibinfo {year} {2016})}\BibitemShut {NoStop}%
	\bibitem [{\citenamefont {de~Leon}\ \emph {et~al.}(2021)\citenamefont
		{de~Leon}, \citenamefont {Itoh}, \citenamefont {Kim}, \citenamefont {Mehta},
		\citenamefont {Northup}, \citenamefont {Paik}, \citenamefont {Palmer},
		\citenamefont {Samarth}, \citenamefont {Sangtawesin},\ and\ \citenamefont
		{Steuerman}}]{DeLeon2021}%
	\BibitemOpen
	\bibfield  {author} {\bibinfo {author} {\bibfnamefont {N.~P.}\ \bibnamefont
			{de~Leon}}, \bibinfo {author} {\bibfnamefont {K.~M.}\ \bibnamefont {Itoh}},
		\bibinfo {author} {\bibfnamefont {D.}~\bibnamefont {Kim}}, \bibinfo {author}
		{\bibfnamefont {K.~K.}\ \bibnamefont {Mehta}}, \bibinfo {author}
		{\bibfnamefont {T.~E.}\ \bibnamefont {Northup}}, \bibinfo {author}
		{\bibfnamefont {H.}~\bibnamefont {Paik}}, \bibinfo {author} {\bibfnamefont
			{B.~S.}\ \bibnamefont {Palmer}}, \bibinfo {author} {\bibfnamefont
			{N.}~\bibnamefont {Samarth}}, \bibinfo {author} {\bibfnamefont
			{S.}~\bibnamefont {Sangtawesin}},\ and\ \bibinfo {author} {\bibfnamefont
			{D.~W.}\ \bibnamefont {Steuerman}},\ }\bibfield  {title} {\enquote {\bibinfo
			{title} {{Materials challenges and opportunities for quantum computing
					hardware}},}\ }\href@noop {} {\bibfield  {journal} {\bibinfo  {journal}
			{Science}\ }\textbf {\bibinfo {volume} {372}},\ \bibinfo {pages} {253}
		(\bibinfo {year} {2021})}\BibitemShut {NoStop}%
	\bibitem [{\citenamefont {Hanson}\ \emph {et~al.}(2007)\citenamefont {Hanson},
		\citenamefont {Kouwenhoven}, \citenamefont {Petta}, \citenamefont {Tarucha},\
		and\ \citenamefont {Vandersypen}}]{Hanson2007}%
	\BibitemOpen
	\bibfield  {author} {\bibinfo {author} {\bibfnamefont {R.}~\bibnamefont
			{Hanson}}, \bibinfo {author} {\bibfnamefont {L.~P.}\ \bibnamefont
			{Kouwenhoven}}, \bibinfo {author} {\bibfnamefont {J.~R.}\ \bibnamefont
			{Petta}}, \bibinfo {author} {\bibfnamefont {S.}~\bibnamefont {Tarucha}},\
		and\ \bibinfo {author} {\bibfnamefont {L.~M.}\ \bibnamefont {Vandersypen}},\
	}\bibfield  {title} {\enquote {\bibinfo {title} {{Spins in few-electron
					quantum dots}},}\ }\href@noop {} {\bibfield  {journal} {\bibinfo  {journal}
			{Rev. Mod. Phys.}\ }\textbf {\bibinfo {volume} {79}},\ \bibinfo {pages}
		{1217--1265} (\bibinfo {year} {2007})}\BibitemShut {NoStop}%
	\bibitem [{\citenamefont {Zwanenburg}\ \emph {et~al.}(2013)\citenamefont
		{Zwanenburg}, \citenamefont {Dzurak}, \citenamefont {Morello}, \citenamefont
		{Simmons}, \citenamefont {Hollenberg}, \citenamefont {Klimeck}, \citenamefont
		{Rogge}, \citenamefont {Coppersmith},\ and\ \citenamefont
		{Eriksson}}]{Zwanenburg2013}%
	\BibitemOpen
	\bibfield  {author} {\bibinfo {author} {\bibfnamefont {F.~A.}\ \bibnamefont
			{Zwanenburg}}, \bibinfo {author} {\bibfnamefont {A.~S.}\ \bibnamefont
			{Dzurak}}, \bibinfo {author} {\bibfnamefont {A.}~\bibnamefont {Morello}},
		\bibinfo {author} {\bibfnamefont {M.~Y.}\ \bibnamefont {Simmons}}, \bibinfo
		{author} {\bibfnamefont {L.~C.}\ \bibnamefont {Hollenberg}}, \bibinfo
		{author} {\bibfnamefont {G.}~\bibnamefont {Klimeck}}, \bibinfo {author}
		{\bibfnamefont {S.}~\bibnamefont {Rogge}}, \bibinfo {author} {\bibfnamefont
			{S.~N.}\ \bibnamefont {Coppersmith}},\ and\ \bibinfo {author} {\bibfnamefont
			{M.~A.}\ \bibnamefont {Eriksson}},\ }\bibfield  {title} {\enquote {\bibinfo
			{title} {{Silicon quantum electronics}},}\ }\href@noop {} {\bibfield
		{journal} {\bibinfo  {journal} {Rev. Mod. Phys.}\ }\textbf {\bibinfo {volume}
			{85}},\ \bibinfo {pages} {961--1019} (\bibinfo {year} {2013})}\BibitemShut
	{NoStop}%
	\bibitem [{\citenamefont {Cole}\ and\ \citenamefont {Cohen}(1969)}]{Cole1969}%
	\BibitemOpen
	\bibfield  {author} {\bibinfo {author} {\bibfnamefont {M.~W.}\ \bibnamefont
			{Cole}}\ and\ \bibinfo {author} {\bibfnamefont {M.~H.}\ \bibnamefont
			{Cohen}},\ }\bibfield  {title} {\enquote {\bibinfo {title}
			{{Image-potential-induced Surface Bands in Insulators}},}\ }\href@noop {}
	{\bibfield  {journal} {\bibinfo  {journal} {Phys. Rev. Lett.}\ }\textbf
		{\bibinfo {volume} {23}},\ \bibinfo {pages} {1238} (\bibinfo {year}
		{1969})}\BibitemShut {NoStop}%
	\bibitem [{\citenamefont {Cole}(1971)}]{Cole1971}%
	\BibitemOpen
	\bibfield  {author} {\bibinfo {author} {\bibfnamefont {M.~W.}\ \bibnamefont
			{Cole}},\ }\bibfield  {title} {\enquote {\bibinfo {title} {{Electronic
					surface states of a dielectric film on a metal substrate}},}\ }\href@noop {}
	{\bibfield  {journal} {\bibinfo  {journal} {Phys. Rev. B}\ }\textbf {\bibinfo
			{volume} {3}},\ \bibinfo {pages} {4418} (\bibinfo {year} {1971})}\BibitemShut
	{NoStop}%
	\bibitem [{\citenamefont {Leiderer}(1992)}]{leiderer1992electrons}%
	\BibitemOpen
	\bibfield  {author} {\bibinfo {author} {\bibfnamefont {P.}~\bibnamefont
			{Leiderer}},\ }\bibfield  {title} {\enquote {\bibinfo {title} {Electrons at
				the surface of quantum systems},}\ }\href@noop {} {\bibfield  {journal}
		{\bibinfo  {journal} {J. Low Temp. Phys.}\ }\textbf {\bibinfo {volume}
			{87}},\ \bibinfo {pages} {247--278} (\bibinfo {year} {1992})}\BibitemShut
	{NoStop}%
	\bibitem [{\citenamefont {Platzman}\ and\ \citenamefont
		{Dykman}(1999)}]{Platzman1999}%
	\BibitemOpen
	\bibfield  {author} {\bibinfo {author} {\bibfnamefont {P.}~\bibnamefont
			{Platzman}}\ and\ \bibinfo {author} {\bibfnamefont {M.~I.}\ \bibnamefont
			{Dykman}},\ }\bibfield  {title} {\enquote {\bibinfo {title} {{Quantum
					computing with electrons on liquid helium}},}\ }\href@noop {} {\bibfield
		{journal} {\bibinfo  {journal} {Science}\ }\textbf {\bibinfo {volume}
			{284}},\ \bibinfo {pages} {1967--1969} (\bibinfo {year} {1999})}\BibitemShut
	{NoStop}%
	\bibitem [{\citenamefont {Smolyaninov}(2001)}]{smolyaninov2001electrons}%
	\BibitemOpen
	\bibfield  {author} {\bibinfo {author} {\bibfnamefont {I.~I.}\ \bibnamefont
			{Smolyaninov}},\ }\bibfield  {title} {\enquote {\bibinfo {title} {Electrons
				on solid hydrogen and solid neon surfaces},}\ }\href@noop {} {\bibfield
		{journal} {\bibinfo  {journal} {Int. J. Mod. Phys. B}\ }\textbf {\bibinfo
			{volume} {15}},\ \bibinfo {pages} {2075--2106} (\bibinfo {year}
		{2001})}\BibitemShut {NoStop}%
	\bibitem [{\citenamefont {Dykman}, \citenamefont {Platzman},\ and\
		\citenamefont {Seddighrad}(2003)}]{Dykman2003}%
	\BibitemOpen
	\bibfield  {author} {\bibinfo {author} {\bibfnamefont {M.~I.}\ \bibnamefont
			{Dykman}}, \bibinfo {author} {\bibfnamefont {P.~M.}\ \bibnamefont
			{Platzman}},\ and\ \bibinfo {author} {\bibfnamefont {P.}~\bibnamefont
			{Seddighrad}},\ }\bibfield  {title} {\enquote {\bibinfo {title} {{Qubits with
					electrons on liquid helium}},}\ }\href@noop {} {\bibfield  {journal}
		{\bibinfo  {journal} {Phys. Rev. B}\ }\textbf {\bibinfo {volume} {67}},\
		\bibinfo {pages} {155402} (\bibinfo {year} {2003})}\BibitemShut {NoStop}%
	\bibitem [{\citenamefont {Lyon}(2006)}]{Lyon2006}%
	\BibitemOpen
	\bibfield  {author} {\bibinfo {author} {\bibfnamefont {S.~A.}\ \bibnamefont
			{Lyon}},\ }\bibfield  {title} {\enquote {\bibinfo {title} {{Spin-based
					quantum computing using electrons on liquid helium}},}\ }\href@noop {}
	{\bibfield  {journal} {\bibinfo  {journal} {Phys. Rev. A}\ }\textbf {\bibinfo
			{volume} {74}},\ \bibinfo {pages} {052338} (\bibinfo {year}
		{2006})}\BibitemShut {NoStop}%
	\bibitem [{\citenamefont {Bradbury}\ \emph {et~al.}(2011)\citenamefont
		{Bradbury}, \citenamefont {Takita}, \citenamefont {Gurrieri}, \citenamefont
		{Wilkel}, \citenamefont {Eng}, \citenamefont {Carroll},\ and\ \citenamefont
		{Lyon}}]{Bradbury2011}%
	\BibitemOpen
	\bibfield  {author} {\bibinfo {author} {\bibfnamefont {F.~R.}\ \bibnamefont
			{Bradbury}}, \bibinfo {author} {\bibfnamefont {M.}~\bibnamefont {Takita}},
		\bibinfo {author} {\bibfnamefont {T.~M.}\ \bibnamefont {Gurrieri}}, \bibinfo
		{author} {\bibfnamefont {K.~J.}\ \bibnamefont {Wilkel}}, \bibinfo {author}
		{\bibfnamefont {K.}~\bibnamefont {Eng}}, \bibinfo {author} {\bibfnamefont
			{M.~S.}\ \bibnamefont {Carroll}},\ and\ \bibinfo {author} {\bibfnamefont
			{S.~A.}\ \bibnamefont {Lyon}},\ }\bibfield  {title} {\enquote {\bibinfo
			{title} {{Efficient clocked electron transfer on superfluid helium}},}\
	}\href@noop {} {\bibfield  {journal} {\bibinfo  {journal} {Phys. Rev. Lett.}\
		}\textbf {\bibinfo {volume} {107}},\ \bibinfo {pages} {266803} (\bibinfo
		{year} {2011})}\BibitemShut {NoStop}%
	\bibitem [{\citenamefont {Wallraff}\ \emph {et~al.}(2004)\citenamefont
		{Wallraff}, \citenamefont {Schuster}, \citenamefont {Blais}, \citenamefont
		{Frunzio}, \citenamefont {Huang}, \citenamefont {Majer}, \citenamefont
		{Kumar}, \citenamefont {Girvin},\ and\ \citenamefont
		{Schoelkopf}}]{Wallraff2004}%
	\BibitemOpen
	\bibfield  {author} {\bibinfo {author} {\bibfnamefont {A.}~\bibnamefont
			{Wallraff}}, \bibinfo {author} {\bibfnamefont {D.~I.}\ \bibnamefont
			{Schuster}}, \bibinfo {author} {\bibfnamefont {A.}~\bibnamefont {Blais}},
		\bibinfo {author} {\bibfnamefont {L.}~\bibnamefont {Frunzio}}, \bibinfo
		{author} {\bibfnamefont {R.~S.}\ \bibnamefont {Huang}}, \bibinfo {author}
		{\bibfnamefont {J.}~\bibnamefont {Majer}}, \bibinfo {author} {\bibfnamefont
			{S.}~\bibnamefont {Kumar}}, \bibinfo {author} {\bibfnamefont {S.~M.}\
			\bibnamefont {Girvin}},\ and\ \bibinfo {author} {\bibfnamefont {R.~J.}\
			\bibnamefont {Schoelkopf}},\ }\bibfield  {title} {\enquote {\bibinfo {title}
			{{Strong coupling of a single photon to a superconducting qubit using circuit
					quantum electrodynamics}},}\ }\href@noop {} {\bibfield  {journal} {\bibinfo
			{journal} {Nature}\ }\textbf {\bibinfo {volume} {431}},\ \bibinfo {pages}
		{162--167} (\bibinfo {year} {2004})}\BibitemShut {NoStop}%
	\bibitem [{\citenamefont {Blais}, \citenamefont {Grimsmo},\ and\ \citenamefont
		{Wallraff}(2021)}]{Blais2021}%
	\BibitemOpen
	\bibfield  {author} {\bibinfo {author} {\bibfnamefont {A.}~\bibnamefont
			{Blais}}, \bibinfo {author} {\bibfnamefont {A.~L.}\ \bibnamefont {Grimsmo}},\
		and\ \bibinfo {author} {\bibfnamefont {A.}~\bibnamefont {Wallraff}},\
	}\bibfield  {title} {\enquote {\bibinfo {title} {{Circuit quantum
					electrodynamics}},}\ }\href@noop {} {\bibfield  {journal} {\bibinfo
			{journal} {Rev. Mod. Phys.}\ }\textbf {\bibinfo {volume} {93}},\ \bibinfo
		{pages} {025005} (\bibinfo {year} {2021})}\BibitemShut {NoStop}%
	\bibitem [{\citenamefont {Schuster}\ \emph {et~al.}(2010)\citenamefont
		{Schuster}, \citenamefont {Fragner}, \citenamefont {Dykman}, \citenamefont
		{Lyon},\ and\ \citenamefont {Schoelkopf}}]{Schuster2010}%
	\BibitemOpen
	\bibfield  {author} {\bibinfo {author} {\bibfnamefont {D.~I.}\ \bibnamefont
			{Schuster}}, \bibinfo {author} {\bibfnamefont {A.}~\bibnamefont {Fragner}},
		\bibinfo {author} {\bibfnamefont {M.~I.}\ \bibnamefont {Dykman}}, \bibinfo
		{author} {\bibfnamefont {S.~A.}\ \bibnamefont {Lyon}},\ and\ \bibinfo
		{author} {\bibfnamefont {R.~J.}\ \bibnamefont {Schoelkopf}},\ }\bibfield
	{title} {\enquote {\bibinfo {title} {{Proposal for manipulating and detecting
					spin and orbital states of trapped electrons on helium using cavity quantum
					electrodynamics}},}\ }\href@noop {} {\bibfield  {journal} {\bibinfo
			{journal} {Phys. Rev. Lett.}\ }\textbf {\bibinfo {volume} {105}},\ \bibinfo
		{pages} {040503} (\bibinfo {year} {2010})}\BibitemShut {NoStop}%
	\bibitem [{\citenamefont {Yang}\ \emph {et~al.}(2016)\citenamefont {Yang},
		\citenamefont {Fragner}, \citenamefont {Koolstra}, \citenamefont {Ocola},
		\citenamefont {Czaplewski}, \citenamefont {Schoelkopf},\ and\ \citenamefont
		{Schuster}}]{Yang2016}%
	\BibitemOpen
	\bibfield  {author} {\bibinfo {author} {\bibfnamefont {G.}~\bibnamefont
			{Yang}}, \bibinfo {author} {\bibfnamefont {A.}~\bibnamefont {Fragner}},
		\bibinfo {author} {\bibfnamefont {G.}~\bibnamefont {Koolstra}}, \bibinfo
		{author} {\bibfnamefont {L.}~\bibnamefont {Ocola}}, \bibinfo {author}
		{\bibfnamefont {D.~A.}\ \bibnamefont {Czaplewski}}, \bibinfo {author}
		{\bibfnamefont {R.~J.}\ \bibnamefont {Schoelkopf}},\ and\ \bibinfo {author}
		{\bibfnamefont {D.~I.}\ \bibnamefont {Schuster}},\ }\bibfield  {title}
	{\enquote {\bibinfo {title} {{Coupling an ensemble of electrons on superfluid
					helium to a superconducting circuit}},}\ }\href@noop {} {\bibfield  {journal}
		{\bibinfo  {journal} {Phys. Rev. X}\ }\textbf {\bibinfo {volume} {6}},\
		\bibinfo {pages} {011031} (\bibinfo {year} {2016})}\BibitemShut {NoStop}%
	\bibitem [{\citenamefont {Koolstra}, \citenamefont {Yang},\ and\ \citenamefont
		{Schuster}(2019)}]{Koolstra2019}%
	\BibitemOpen
	\bibfield  {author} {\bibinfo {author} {\bibfnamefont {G.}~\bibnamefont
			{Koolstra}}, \bibinfo {author} {\bibfnamefont {G.}~\bibnamefont {Yang}},\
		and\ \bibinfo {author} {\bibfnamefont {D.~I.}\ \bibnamefont {Schuster}},\
	}\bibfield  {title} {\enquote {\bibinfo {title} {{Coupling a single electron
					on superfluid helium to a superconducting resonator}},}\ }\href@noop {}
	{\bibfield  {journal} {\bibinfo  {journal} {Nat. Commun.}\ }\textbf {\bibinfo
			{volume} {10}},\ \bibinfo {pages} {5323} (\bibinfo {year}
		{2019})}\BibitemShut {NoStop}%
	\bibitem [{\citenamefont {Jin}(2020)}]{jin2020quantum}%
	\BibitemOpen
	\bibfield  {author} {\bibinfo {author} {\bibfnamefont {D.}~\bibnamefont
			{Jin}},\ }\bibfield  {title} {\enquote {\bibinfo {title} {Quantum electronics
				and optics at the interface of solid neon and superfluid helium},}\
	}\href@noop {} {\bibfield  {journal} {\bibinfo  {journal} {Quantum Sci.
				Technol.}\ }\textbf {\bibinfo {volume} {5}},\ \bibinfo {pages} {035003}
		(\bibinfo {year} {2020})}\BibitemShut {NoStop}%
	\bibitem [{\citenamefont {Clerk}\ \emph {et~al.}(2020)\citenamefont {Clerk},
		\citenamefont {Lehnert}, \citenamefont {Bertet}, \citenamefont {Petta},\ and\
		\citenamefont {Nakamura}}]{Clerk2020}%
	\BibitemOpen
	\bibfield  {author} {\bibinfo {author} {\bibfnamefont {A.~A.}\ \bibnamefont
			{Clerk}}, \bibinfo {author} {\bibfnamefont {K.~W.}\ \bibnamefont {Lehnert}},
		\bibinfo {author} {\bibfnamefont {P.}~\bibnamefont {Bertet}}, \bibinfo
		{author} {\bibfnamefont {J.~R.}\ \bibnamefont {Petta}},\ and\ \bibinfo
		{author} {\bibfnamefont {Y.}~\bibnamefont {Nakamura}},\ }\bibfield  {title}
	{\enquote {\bibinfo {title} {{Hybrid quantum systems with circuit quantum
					electrodynamics}},}\ }\href@noop {} {\bibfield  {journal} {\bibinfo
			{journal} {Nat. Phys.}\ }\textbf {\bibinfo {volume} {16}},\ \bibinfo {pages}
		{257--267} (\bibinfo {year} {2020})}\BibitemShut {NoStop}%
	\bibitem [{\citenamefont {Chatterjee}\ \emph {et~al.}(2021)\citenamefont
		{Chatterjee}, \citenamefont {Stevenson}, \citenamefont {{De Franceschi}},
		\citenamefont {Morello}, \citenamefont {de~Leon},\ and\ \citenamefont
		{Kuemmeth}}]{Chatterjee2021}%
	\BibitemOpen
	\bibfield  {author} {\bibinfo {author} {\bibfnamefont {A.}~\bibnamefont
			{Chatterjee}}, \bibinfo {author} {\bibfnamefont {P.}~\bibnamefont
			{Stevenson}}, \bibinfo {author} {\bibfnamefont {S.}~\bibnamefont {{De
					Franceschi}}}, \bibinfo {author} {\bibfnamefont {A.}~\bibnamefont {Morello}},
		\bibinfo {author} {\bibfnamefont {N.~P.}\ \bibnamefont {de~Leon}},\ and\
		\bibinfo {author} {\bibfnamefont {F.}~\bibnamefont {Kuemmeth}},\ }\bibfield
	{title} {\enquote {\bibinfo {title} {{Semiconductor qubits in practice}},}\
	}\href@noop {} {\bibfield  {journal} {\bibinfo  {journal} {Nat. Rev. Phys.}\
		}\textbf {\bibinfo {volume} {3}},\ \bibinfo {pages} {157--177} (\bibinfo
		{year} {2021})}\BibitemShut {NoStop}%
	\bibitem [{\citenamefont {Nakamura}, \citenamefont {Pashkin},\ and\
		\citenamefont {Tsai}(1999)}]{nakamura1999coherent}%
	\BibitemOpen
	\bibfield  {author} {\bibinfo {author} {\bibfnamefont {Y.}~\bibnamefont
			{Nakamura}}, \bibinfo {author} {\bibfnamefont {Y.~A.}\ \bibnamefont
			{Pashkin}},\ and\ \bibinfo {author} {\bibfnamefont {J.~S.}\ \bibnamefont
			{Tsai}},\ }\bibfield  {title} {\enquote {\bibinfo {title} {Coherent control
				of macroscopic quantum states in a single-cooper-pair box},}\ }\href@noop {}
	{\bibfield  {journal} {\bibinfo  {journal} {Nature}\ }\textbf {\bibinfo
			{volume} {398}},\ \bibinfo {pages} {786--788} (\bibinfo {year}
		{1999})}\BibitemShut {NoStop}%
	\bibitem [{\citenamefont {Schoelkopf}\ and\ \citenamefont
		{Girvin}(2008)}]{Schoelkopf2008}%
	\BibitemOpen
	\bibfield  {author} {\bibinfo {author} {\bibfnamefont {R.~J.}\ \bibnamefont
			{Schoelkopf}}\ and\ \bibinfo {author} {\bibfnamefont {S.~M.}\ \bibnamefont
			{Girvin}},\ }\bibfield  {title} {\enquote {\bibinfo {title} {{Wiring up
					quantum systems}},}\ }\href@noop {} {\bibfield  {journal} {\bibinfo
			{journal} {Nature}\ }\textbf {\bibinfo {volume} {451}},\ \bibinfo {pages}
		{664--669} (\bibinfo {year} {2008})}\BibitemShut {NoStop}%
	\bibitem [{\citenamefont {Clarke}\ and\ \citenamefont
		{Wilhelm}(2008)}]{Clarke2008}%
	\BibitemOpen
	\bibfield  {author} {\bibinfo {author} {\bibfnamefont {J.}~\bibnamefont
			{Clarke}}\ and\ \bibinfo {author} {\bibfnamefont {F.~K.}\ \bibnamefont
			{Wilhelm}},\ }\bibfield  {title} {\enquote {\bibinfo {title}
			{{Superconducting quantum bits}},}\ }\href@noop {} {\bibfield  {journal}
		{\bibinfo  {journal} {Nature}\ }\textbf {\bibinfo {volume} {453}},\ \bibinfo
		{pages} {1031--1042} (\bibinfo {year} {2008})}\BibitemShut {NoStop}%
	\bibitem [{\citenamefont {Arute}\ \emph {et~al.}(2019)\citenamefont {Arute},
		\citenamefont {Arya}, \citenamefont {Babbush} \emph {et~al.}}]{Arute2019}%
	\BibitemOpen
	\bibfield  {author} {\bibinfo {author} {\bibfnamefont {F.}~\bibnamefont
			{Arute}}, \bibinfo {author} {\bibfnamefont {K.}~\bibnamefont {Arya}},
		\bibinfo {author} {\bibfnamefont {R.}~\bibnamefont {Babbush}}, \emph
		{et~al.},\ }\bibfield  {title} {\enquote {\bibinfo {title} {{Quantum
					supremacy using a programmable superconducting processor}},}\ }\href@noop {}
	{\bibfield  {journal} {\bibinfo  {journal} {Nature}\ }\textbf {\bibinfo
			{volume} {574}},\ \bibinfo {pages} {505--510} (\bibinfo {year}
		{2019})}\BibitemShut {NoStop}%
	\bibitem [{\citenamefont {Mi}\ \emph {et~al.}(2017)\citenamefont {Mi},
		\citenamefont {Cady}, \citenamefont {Zajac}, \citenamefont {Deelman},\ and\
		\citenamefont {Petta}}]{Mi2017}%
	\BibitemOpen
	\bibfield  {author} {\bibinfo {author} {\bibfnamefont {X.}~\bibnamefont
			{Mi}}, \bibinfo {author} {\bibfnamefont {J.~V.}\ \bibnamefont {Cady}},
		\bibinfo {author} {\bibfnamefont {D.~M.}\ \bibnamefont {Zajac}}, \bibinfo
		{author} {\bibfnamefont {P.~W.}\ \bibnamefont {Deelman}},\ and\ \bibinfo
		{author} {\bibfnamefont {J.~R.}\ \bibnamefont {Petta}},\ }\bibfield  {title}
	{\enquote {\bibinfo {title} {{Strong coupling of a single electron in silicon
					to a microwave photon}},}\ }\href@noop {} {\bibfield  {journal} {\bibinfo
			{journal} {Science}\ }\textbf {\bibinfo {volume} {355}},\ \bibinfo {pages}
		{156--158} (\bibinfo {year} {2017})}\BibitemShut {NoStop}%
	\bibitem [{\citenamefont {Mi}\ \emph {et~al.}(2018)\citenamefont {Mi},
		\citenamefont {Benito}, \citenamefont {Putz}, \citenamefont {Zajac},
		\citenamefont {Taylor}, \citenamefont {Burkard},\ and\ \citenamefont
		{Petta}}]{Mi2018}%
	\BibitemOpen
	\bibfield  {author} {\bibinfo {author} {\bibfnamefont {X.}~\bibnamefont
			{Mi}}, \bibinfo {author} {\bibfnamefont {M.}~\bibnamefont {Benito}}, \bibinfo
		{author} {\bibfnamefont {S.}~\bibnamefont {Putz}}, \bibinfo {author}
		{\bibfnamefont {D.~M.}\ \bibnamefont {Zajac}}, \bibinfo {author}
		{\bibfnamefont {J.~M.}\ \bibnamefont {Taylor}}, \bibinfo {author}
		{\bibfnamefont {G.}~\bibnamefont {Burkard}},\ and\ \bibinfo {author}
		{\bibfnamefont {J.~R.}\ \bibnamefont {Petta}},\ }\bibfield  {title} {\enquote
		{\bibinfo {title} {{A coherent spin-photon interface in silicon}},}\
	}\href@noop {} {\bibfield  {journal} {\bibinfo  {journal} {Nature}\ }\textbf
		{\bibinfo {volume} {555}},\ \bibinfo {pages} {599--603} (\bibinfo {year}
		{2018})}\BibitemShut {NoStop}%
	\bibitem [{\citenamefont {Samkharadze}\ \emph {et~al.}(2018)\citenamefont
		{Samkharadze}, \citenamefont {Zheng}, \citenamefont {Kalhor}, \citenamefont
		{Brousse}, \citenamefont {Sammak}, \citenamefont {Mendes}, \citenamefont
		{Blais}, \citenamefont {Scappucci},\ and\ \citenamefont
		{Vandersypen}}]{Samkharadze2018}%
	\BibitemOpen
	\bibfield  {author} {\bibinfo {author} {\bibfnamefont {N.}~\bibnamefont
			{Samkharadze}}, \bibinfo {author} {\bibfnamefont {G.}~\bibnamefont {Zheng}},
		\bibinfo {author} {\bibfnamefont {N.}~\bibnamefont {Kalhor}}, \bibinfo
		{author} {\bibfnamefont {D.}~\bibnamefont {Brousse}}, \bibinfo {author}
		{\bibfnamefont {A.}~\bibnamefont {Sammak}}, \bibinfo {author} {\bibfnamefont
			{U.~C.}\ \bibnamefont {Mendes}}, \bibinfo {author} {\bibfnamefont
			{A.}~\bibnamefont {Blais}}, \bibinfo {author} {\bibfnamefont
			{G.}~\bibnamefont {Scappucci}},\ and\ \bibinfo {author} {\bibfnamefont
			{L.~M.}\ \bibnamefont {Vandersypen}},\ }\bibfield  {title} {\enquote
		{\bibinfo {title} {{Strong spin-photon coupling in silicon}},}\ }\href@noop
	{} {\bibfield  {journal} {\bibinfo  {journal} {Science}\ }\textbf {\bibinfo
			{volume} {359}},\ \bibinfo {pages} {1123--1127} (\bibinfo {year}
		{2018})}\BibitemShut {NoStop}%
	\bibitem [{\citenamefont {Landig}\ \emph {et~al.}(2018)\citenamefont {Landig},
		\citenamefont {Koski}, \citenamefont {Scarlino}, \citenamefont {Mendes},
		\citenamefont {Blais}, \citenamefont {Reichl}, \citenamefont {Wegscheider},
		\citenamefont {Wallraff}, \citenamefont {Ensslin},\ and\ \citenamefont
		{Ihn}}]{Landig2018}%
	\BibitemOpen
	\bibfield  {author} {\bibinfo {author} {\bibfnamefont {A.~J.}\ \bibnamefont
			{Landig}}, \bibinfo {author} {\bibfnamefont {J.~V.}\ \bibnamefont {Koski}},
		\bibinfo {author} {\bibfnamefont {P.}~\bibnamefont {Scarlino}}, \bibinfo
		{author} {\bibfnamefont {U.~C.}\ \bibnamefont {Mendes}}, \bibinfo {author}
		{\bibfnamefont {A.}~\bibnamefont {Blais}}, \bibinfo {author} {\bibfnamefont
			{C.}~\bibnamefont {Reichl}}, \bibinfo {author} {\bibfnamefont
			{W.}~\bibnamefont {Wegscheider}}, \bibinfo {author} {\bibfnamefont
			{A.}~\bibnamefont {Wallraff}}, \bibinfo {author} {\bibfnamefont
			{K.}~\bibnamefont {Ensslin}},\ and\ \bibinfo {author} {\bibfnamefont
			{T.}~\bibnamefont {Ihn}},\ }\bibfield  {title} {\enquote {\bibinfo {title}
			{{Coherent spin–photon coupling using a resonant exchange qubit}},}\
	}\href@noop {} {\bibfield  {journal} {\bibinfo  {journal} {Nature}\ }\textbf
		{\bibinfo {volume} {560}},\ \bibinfo {pages} {179--184} (\bibinfo {year}
		{2018})}\BibitemShut {NoStop}%
	\bibitem [{\citenamefont {Petit}\ \emph {et~al.}(2020)\citenamefont {Petit},
		\citenamefont {Eenink}, \citenamefont {Russ}, \citenamefont {Lawrie},
		\citenamefont {Hendrickx}, \citenamefont {Philips}, \citenamefont {Clarke},
		\citenamefont {Vandersypen},\ and\ \citenamefont
		{Veldhorst}}]{petit2020universal}%
	\BibitemOpen
	\bibfield  {author} {\bibinfo {author} {\bibfnamefont {L.}~\bibnamefont
			{Petit}}, \bibinfo {author} {\bibfnamefont {H.}~\bibnamefont {Eenink}},
		\bibinfo {author} {\bibfnamefont {M.}~\bibnamefont {Russ}}, \bibinfo {author}
		{\bibfnamefont {W.}~\bibnamefont {Lawrie}}, \bibinfo {author} {\bibfnamefont
			{N.}~\bibnamefont {Hendrickx}}, \bibinfo {author} {\bibfnamefont
			{S.}~\bibnamefont {Philips}}, \bibinfo {author} {\bibfnamefont
			{J.}~\bibnamefont {Clarke}}, \bibinfo {author} {\bibfnamefont
			{L.}~\bibnamefont {Vandersypen}},\ and\ \bibinfo {author} {\bibfnamefont
			{M.}~\bibnamefont {Veldhorst}},\ }\bibfield  {title} {\enquote {\bibinfo
			{title} {Universal quantum logic in hot silicon qubits},}\ }\href@noop {}
	{\bibfield  {journal} {\bibinfo  {journal} {Nature}\ }\textbf {\bibinfo
			{volume} {580}},\ \bibinfo {pages} {355--359} (\bibinfo {year}
		{2020})}\BibitemShut {NoStop}%
	\bibitem [{\citenamefont {Burkard}\ \emph {et~al.}(2020)\citenamefont
		{Burkard}, \citenamefont {Gullans}, \citenamefont {Mi},\ and\ \citenamefont
		{Petta}}]{Burkard2020}%
	\BibitemOpen
	\bibfield  {author} {\bibinfo {author} {\bibfnamefont {G.}~\bibnamefont
			{Burkard}}, \bibinfo {author} {\bibfnamefont {M.~J.}\ \bibnamefont
			{Gullans}}, \bibinfo {author} {\bibfnamefont {X.}~\bibnamefont {Mi}},\ and\
		\bibinfo {author} {\bibfnamefont {J.~R.}\ \bibnamefont {Petta}},\ }\bibfield
	{title} {\enquote {\bibinfo {title} {{Superconductor-semiconductor
					hybrid-circuit quantum electrodynamics}},}\ }\href@noop {} {\bibfield
		{journal} {\bibinfo  {journal} {Nat. Rev. Phys.}\ }\textbf {\bibinfo {volume}
			{2}},\ \bibinfo {pages} {129--140} (\bibinfo {year} {2020})}\BibitemShut
	{NoStop}%
	\bibitem [{\citenamefont {Monroe}\ \emph {et~al.}(1995)\citenamefont {Monroe},
		\citenamefont {Meekhof}, \citenamefont {King}, \citenamefont {Itano},\ and\
		\citenamefont {Wineland}}]{monroe1995demonstration}%
	\BibitemOpen
	\bibfield  {author} {\bibinfo {author} {\bibfnamefont {C.}~\bibnamefont
			{Monroe}}, \bibinfo {author} {\bibfnamefont {D.~M.}\ \bibnamefont {Meekhof}},
		\bibinfo {author} {\bibfnamefont {B.~E.}\ \bibnamefont {King}}, \bibinfo
		{author} {\bibfnamefont {W.~M.}\ \bibnamefont {Itano}},\ and\ \bibinfo
		{author} {\bibfnamefont {D.~J.}\ \bibnamefont {Wineland}},\ }\bibfield
	{title} {\enquote {\bibinfo {title} {Demonstration of a fundamental quantum
				logic gate},}\ }\href@noop {} {\bibfield  {journal} {\bibinfo  {journal}
			{Phys. Rev. Lett.}\ }\textbf {\bibinfo {volume} {75}},\ \bibinfo {pages}
		{4714} (\bibinfo {year} {1995})}\BibitemShut {NoStop}%
	\bibitem [{\citenamefont {Kielpinski}, \citenamefont {Monroe},\ and\
		\citenamefont {Wineland}(2002)}]{kielpinski2002}%
	\BibitemOpen
	\bibfield  {author} {\bibinfo {author} {\bibfnamefont {D.}~\bibnamefont
			{Kielpinski}}, \bibinfo {author} {\bibfnamefont {C.}~\bibnamefont {Monroe}},\
		and\ \bibinfo {author} {\bibfnamefont {D.~J.}\ \bibnamefont {Wineland}},\
	}\bibfield  {title} {\enquote {\bibinfo {title} {Architecture for a
				large-scale ion-trap quantum computer},}\ }\href@noop {} {\bibfield
		{journal} {\bibinfo  {journal} {Nature}\ }\textbf {\bibinfo {volume} {417}},\
		\bibinfo {pages} {709--711} (\bibinfo {year} {2002})}\BibitemShut {NoStop}%
	\bibitem [{\citenamefont {Leibfried}\ \emph {et~al.}(2003)\citenamefont
		{Leibfried}, \citenamefont {Blatt}, \citenamefont {Monroe},\ and\
		\citenamefont {Wineland}}]{leibfried2003quantum}%
	\BibitemOpen
	\bibfield  {author} {\bibinfo {author} {\bibfnamefont {D.}~\bibnamefont
			{Leibfried}}, \bibinfo {author} {\bibfnamefont {R.}~\bibnamefont {Blatt}},
		\bibinfo {author} {\bibfnamefont {C.}~\bibnamefont {Monroe}},\ and\ \bibinfo
		{author} {\bibfnamefont {D.}~\bibnamefont {Wineland}},\ }\bibfield  {title}
	{\enquote {\bibinfo {title} {Quantum dynamics of single trapped ions},}\
	}\href@noop {} {\bibfield  {journal} {\bibinfo  {journal} {Rev. Mod. Phys.}\
		}\textbf {\bibinfo {volume} {75}},\ \bibinfo {pages} {281} (\bibinfo {year}
		{2003})}\BibitemShut {NoStop}%
	\bibitem [{\citenamefont {Bruzewicz}\ \emph {et~al.}(2019)\citenamefont
		{Bruzewicz}, \citenamefont {Chiaverini}, \citenamefont {McConnell},\ and\
		\citenamefont {Sage}}]{Bruzewicz2019}%
	\BibitemOpen
	\bibfield  {author} {\bibinfo {author} {\bibfnamefont {C.~D.}\ \bibnamefont
			{Bruzewicz}}, \bibinfo {author} {\bibfnamefont {J.}~\bibnamefont
			{Chiaverini}}, \bibinfo {author} {\bibfnamefont {R.}~\bibnamefont
			{McConnell}},\ and\ \bibinfo {author} {\bibfnamefont {J.~M.}\ \bibnamefont
			{Sage}},\ }\bibfield  {title} {\enquote {\bibinfo {title} {{Trapped-ion
					quantum computing: Progress and challenges}},}\ }\href@noop {} {\bibfield
		{journal} {\bibinfo  {journal} {Appl. Phys. Rev.}\ }\textbf {\bibinfo
			{volume} {6}},\ \bibinfo {pages} {021314} (\bibinfo {year}
		{2019})}\BibitemShut {NoStop}%
	\bibitem [{\citenamefont {Pino}\ \emph {et~al.}(2021)\citenamefont {Pino},
		\citenamefont {Dreiling}, \citenamefont {Figgatt}, \citenamefont {Gaebler},
		\citenamefont {Moses}, \citenamefont {Allman}, \citenamefont {Baldwin},
		\citenamefont {Foss-Feig}, \citenamefont {Hayes}, \citenamefont {Mayer},
		\citenamefont {Ryan-Anderson},\ and\ \citenamefont {Neyenhuis}}]{Pino2021}%
	\BibitemOpen
	\bibfield  {author} {\bibinfo {author} {\bibfnamefont {J.~M.}\ \bibnamefont
			{Pino}}, \bibinfo {author} {\bibfnamefont {J.~M.}\ \bibnamefont {Dreiling}},
		\bibinfo {author} {\bibfnamefont {C.}~\bibnamefont {Figgatt}}, \bibinfo
		{author} {\bibfnamefont {J.~P.}\ \bibnamefont {Gaebler}}, \bibinfo {author}
		{\bibfnamefont {S.~A.}\ \bibnamefont {Moses}}, \bibinfo {author}
		{\bibfnamefont {M.~S.}\ \bibnamefont {Allman}}, \bibinfo {author}
		{\bibfnamefont {C.~H.}\ \bibnamefont {Baldwin}}, \bibinfo {author}
		{\bibfnamefont {M.}~\bibnamefont {Foss-Feig}}, \bibinfo {author}
		{\bibfnamefont {D.}~\bibnamefont {Hayes}}, \bibinfo {author} {\bibfnamefont
			{K.}~\bibnamefont {Mayer}}, \bibinfo {author} {\bibfnamefont
			{C.}~\bibnamefont {Ryan-Anderson}},\ and\ \bibinfo {author} {\bibfnamefont
			{B.}~\bibnamefont {Neyenhuis}},\ }\bibfield  {title} {\enquote {\bibinfo
			{title} {{Demonstration of the trapped-ion quantum CCD computer
					architecture}},}\ }\href@noop {} {\bibfield  {journal} {\bibinfo  {journal}
			{Nature}\ }\textbf {\bibinfo {volume} {592}},\ \bibinfo {pages} {209--213}
		(\bibinfo {year} {2021})}\BibitemShut {NoStop}%
	\bibitem [{\citenamefont {Brennen}\ \emph {et~al.}(1999)\citenamefont
		{Brennen}, \citenamefont {Caves}, \citenamefont {Jessen},\ and\ \citenamefont
		{Deutsch}}]{brennen1999quantum}%
	\BibitemOpen
	\bibfield  {author} {\bibinfo {author} {\bibfnamefont {G.~K.}\ \bibnamefont
			{Brennen}}, \bibinfo {author} {\bibfnamefont {C.~M.}\ \bibnamefont {Caves}},
		\bibinfo {author} {\bibfnamefont {P.~S.}\ \bibnamefont {Jessen}},\ and\
		\bibinfo {author} {\bibfnamefont {I.~H.}\ \bibnamefont {Deutsch}},\
	}\bibfield  {title} {\enquote {\bibinfo {title} {Quantum logic gates in
				optical lattices},}\ }\href@noop {} {\bibfield  {journal} {\bibinfo
			{journal} {Phys. Rev. Lett.}\ }\textbf {\bibinfo {volume} {82}},\ \bibinfo
		{pages} {1060} (\bibinfo {year} {1999})}\BibitemShut {NoStop}%
	\bibitem [{\citenamefont {Jaksch}\ \emph {et~al.}(2000)\citenamefont {Jaksch},
		\citenamefont {Cirac}, \citenamefont {Zoller}, \citenamefont {Rolston},
		\citenamefont {C{\^o}t{\'e}},\ and\ \citenamefont {Lukin}}]{jaksch2000fast}%
	\BibitemOpen
	\bibfield  {author} {\bibinfo {author} {\bibfnamefont {D.}~\bibnamefont
			{Jaksch}}, \bibinfo {author} {\bibfnamefont {J.~I.}\ \bibnamefont {Cirac}},
		\bibinfo {author} {\bibfnamefont {P.}~\bibnamefont {Zoller}}, \bibinfo
		{author} {\bibfnamefont {S.~L.}\ \bibnamefont {Rolston}}, \bibinfo {author}
		{\bibfnamefont {R.}~\bibnamefont {C{\^o}t{\'e}}},\ and\ \bibinfo {author}
		{\bibfnamefont {M.~D.}\ \bibnamefont {Lukin}},\ }\bibfield  {title} {\enquote
		{\bibinfo {title} {Fast quantum gates for neutral atoms},}\ }\href@noop {}
	{\bibfield  {journal} {\bibinfo  {journal} {Phys. Rev. Lett.}\ }\textbf
		{\bibinfo {volume} {85}},\ \bibinfo {pages} {2208} (\bibinfo {year}
		{2000})}\BibitemShut {NoStop}%
	\bibitem [{\citenamefont {Saffman}, \citenamefont {Walker},\ and\ \citenamefont
		{M{\o}lmer}(2010)}]{Saffman2010}%
	\BibitemOpen
	\bibfield  {author} {\bibinfo {author} {\bibfnamefont {M.}~\bibnamefont
			{Saffman}}, \bibinfo {author} {\bibfnamefont {T.~G.}\ \bibnamefont
			{Walker}},\ and\ \bibinfo {author} {\bibfnamefont {K.}~\bibnamefont
			{M{\o}lmer}},\ }\bibfield  {title} {\enquote {\bibinfo {title} {{Quantum
					information with Rydberg atoms}},}\ }\href@noop {} {\bibfield  {journal}
		{\bibinfo  {journal} {Rev. Mod. Phys.}\ }\textbf {\bibinfo {volume} {82}},\
		\bibinfo {pages} {2313--2363} (\bibinfo {year} {2010})}\BibitemShut {NoStop}%
	\bibitem [{\citenamefont {Wang}\ \emph {et~al.}(2016)\citenamefont {Wang},
		\citenamefont {Kumar}, \citenamefont {Wu},\ and\ \citenamefont
		{Weiss}}]{wang2016single}%
	\BibitemOpen
	\bibfield  {author} {\bibinfo {author} {\bibfnamefont {Y.}~\bibnamefont
			{Wang}}, \bibinfo {author} {\bibfnamefont {A.}~\bibnamefont {Kumar}},
		\bibinfo {author} {\bibfnamefont {T.-Y.}\ \bibnamefont {Wu}},\ and\ \bibinfo
		{author} {\bibfnamefont {D.~S.}\ \bibnamefont {Weiss}},\ }\bibfield  {title}
	{\enquote {\bibinfo {title} {Single-qubit gates based on targeted phase
				shifts in a {3D} neutral atom array},}\ }\href@noop {} {\bibfield  {journal}
		{\bibinfo  {journal} {Science}\ }\textbf {\bibinfo {volume} {352}},\ \bibinfo
		{pages} {1562--1565} (\bibinfo {year} {2016})}\BibitemShut {NoStop}%
	\bibitem [{\citenamefont {Pla}\ \emph {et~al.}(2012)\citenamefont {Pla},
		\citenamefont {Tan}, \citenamefont {Dehollain}, \citenamefont {Lim},
		\citenamefont {Morton}, \citenamefont {Jamieson}, \citenamefont {Dzurak},\
		and\ \citenamefont {Morello}}]{Pla2012}%
	\BibitemOpen
	\bibfield  {author} {\bibinfo {author} {\bibfnamefont {J.~J.}\ \bibnamefont
			{Pla}}, \bibinfo {author} {\bibfnamefont {K.~Y.}\ \bibnamefont {Tan}},
		\bibinfo {author} {\bibfnamefont {J.~P.}\ \bibnamefont {Dehollain}}, \bibinfo
		{author} {\bibfnamefont {W.~H.}\ \bibnamefont {Lim}}, \bibinfo {author}
		{\bibfnamefont {J.~J.}\ \bibnamefont {Morton}}, \bibinfo {author}
		{\bibfnamefont {D.~N.}\ \bibnamefont {Jamieson}}, \bibinfo {author}
		{\bibfnamefont {A.~S.}\ \bibnamefont {Dzurak}},\ and\ \bibinfo {author}
		{\bibfnamefont {A.}~\bibnamefont {Morello}},\ }\bibfield  {title} {\enquote
		{\bibinfo {title} {{A single-atom electron spin qubit in silicon}},}\
	}\href@noop {} {\bibfield  {journal} {\bibinfo  {journal} {Nature}\ }\textbf
		{\bibinfo {volume} {489}},\ \bibinfo {pages} {541--545} (\bibinfo {year}
		{2012})}\BibitemShut {NoStop}%
	\bibitem [{\citenamefont {Pla}\ \emph {et~al.}(2013)\citenamefont {Pla},
		\citenamefont {Tan}, \citenamefont {Dehollain}, \citenamefont {Lim},
		\citenamefont {Morton}, \citenamefont {Zwanenburg}, \citenamefont {Jamieson},
		\citenamefont {Dzurak},\ and\ \citenamefont {Morello}}]{Pla2013}%
	\BibitemOpen
	\bibfield  {author} {\bibinfo {author} {\bibfnamefont {J.~J.}\ \bibnamefont
			{Pla}}, \bibinfo {author} {\bibfnamefont {K.~Y.}\ \bibnamefont {Tan}},
		\bibinfo {author} {\bibfnamefont {J.~P.}\ \bibnamefont {Dehollain}}, \bibinfo
		{author} {\bibfnamefont {W.~H.}\ \bibnamefont {Lim}}, \bibinfo {author}
		{\bibfnamefont {J.~J.}\ \bibnamefont {Morton}}, \bibinfo {author}
		{\bibfnamefont {F.~A.}\ \bibnamefont {Zwanenburg}}, \bibinfo {author}
		{\bibfnamefont {D.~N.}\ \bibnamefont {Jamieson}}, \bibinfo {author}
		{\bibfnamefont {A.~S.}\ \bibnamefont {Dzurak}},\ and\ \bibinfo {author}
		{\bibfnamefont {A.}~\bibnamefont {Morello}},\ }\bibfield  {title} {\enquote
		{\bibinfo {title} {{High-fidelity readout and control of a nuclear spin qubit
					in silicon}},}\ }\href@noop {} {\bibfield  {journal} {\bibinfo  {journal}
			{Nature}\ }\textbf {\bibinfo {volume} {496}},\ \bibinfo {pages} {334--338}
		(\bibinfo {year} {2013})}\BibitemShut {NoStop}%
	\bibitem [{\citenamefont {Chen}\ \emph {et~al.}(2020)\citenamefont {Chen},
		\citenamefont {Raha}, \citenamefont {Phenicie}, \citenamefont {Ourari},\ and\
		\citenamefont {Thompson}}]{Chen2020}%
	\BibitemOpen
	\bibfield  {author} {\bibinfo {author} {\bibfnamefont {S.}~\bibnamefont
			{Chen}}, \bibinfo {author} {\bibfnamefont {M.}~\bibnamefont {Raha}}, \bibinfo
		{author} {\bibfnamefont {C.~M.}\ \bibnamefont {Phenicie}}, \bibinfo {author}
		{\bibfnamefont {S.}~\bibnamefont {Ourari}},\ and\ \bibinfo {author}
		{\bibfnamefont {J.~D.}\ \bibnamefont {Thompson}},\ }\bibfield  {title}
	{\enquote {\bibinfo {title} {{Parallel single-shot measurement and coherent
					control of solid-state spins below the diffraction limit}},}\ }\href@noop {}
	{\bibfield  {journal} {\bibinfo  {journal} {Science}\ }\textbf {\bibinfo
			{volume} {370}},\ \bibinfo {pages} {592--595} (\bibinfo {year}
		{2020})}\BibitemShut {NoStop}%
	\bibitem [{\citenamefont {Wolfowicz}\ \emph {et~al.}(2021)\citenamefont
		{Wolfowicz}, \citenamefont {Heremans}, \citenamefont {Anderson},
		\citenamefont {Kanai}, \citenamefont {Seo}, \citenamefont {Gali},
		\citenamefont {Galli},\ and\ \citenamefont {Awschalom}}]{Wolfowicz2021}%
	\BibitemOpen
	\bibfield  {author} {\bibinfo {author} {\bibfnamefont {G.}~\bibnamefont
			{Wolfowicz}}, \bibinfo {author} {\bibfnamefont {F.~J.}\ \bibnamefont
			{Heremans}}, \bibinfo {author} {\bibfnamefont {C.~P.}\ \bibnamefont
			{Anderson}}, \bibinfo {author} {\bibfnamefont {S.}~\bibnamefont {Kanai}},
		\bibinfo {author} {\bibfnamefont {H.}~\bibnamefont {Seo}}, \bibinfo {author}
		{\bibfnamefont {A.}~\bibnamefont {Gali}}, \bibinfo {author} {\bibfnamefont
			{G.}~\bibnamefont {Galli}},\ and\ \bibinfo {author} {\bibfnamefont {D.~D.}\
			\bibnamefont {Awschalom}},\ }\bibfield  {title} {\enquote {\bibinfo {title}
			{{Quantum guidelines for solid-state spin defects}},}\ }\href@noop {}
	{\bibfield  {journal} {\bibinfo  {journal} {Nat. Rev. Mater.}\ }\textbf
		{\bibinfo {volume} {6}},\ \bibinfo {pages} {906--925} (\bibinfo {year}
		{2021})}\BibitemShut {NoStop}%
	\bibitem [{\citenamefont {Vincent}\ \emph {et~al.}(2012)\citenamefont
		{Vincent}, \citenamefont {Klyatskaya}, \citenamefont {Ruben}, \citenamefont
		{Wernsdorfer},\ and\ \citenamefont {Balestro}}]{Vincent2012}%
	\BibitemOpen
	\bibfield  {author} {\bibinfo {author} {\bibfnamefont {R.}~\bibnamefont
			{Vincent}}, \bibinfo {author} {\bibfnamefont {S.}~\bibnamefont {Klyatskaya}},
		\bibinfo {author} {\bibfnamefont {M.}~\bibnamefont {Ruben}}, \bibinfo
		{author} {\bibfnamefont {W.}~\bibnamefont {Wernsdorfer}},\ and\ \bibinfo
		{author} {\bibfnamefont {F.}~\bibnamefont {Balestro}},\ }\bibfield  {title}
	{\enquote {\bibinfo {title} {{Electronic read-out of a single nuclear spin
					using a molecular spin transistor}},}\ }\href@noop {} {\bibfield  {journal}
		{\bibinfo  {journal} {Nature}\ }\textbf {\bibinfo {volume} {488}},\ \bibinfo
		{pages} {357--360} (\bibinfo {year} {2012})}\BibitemShut {NoStop}%
	\bibitem [{\citenamefont {Thiele}\ \emph {et~al.}(2014)\citenamefont {Thiele},
		\citenamefont {Balestro}, \citenamefont {Ballou}, \citenamefont {Klyatskaya},
		\citenamefont {Ruben},\ and\ \citenamefont {Wernsdorfer}}]{Thiele2014}%
	\BibitemOpen
	\bibfield  {author} {\bibinfo {author} {\bibfnamefont {S.}~\bibnamefont
			{Thiele}}, \bibinfo {author} {\bibfnamefont {F.}~\bibnamefont {Balestro}},
		\bibinfo {author} {\bibfnamefont {R.}~\bibnamefont {Ballou}}, \bibinfo
		{author} {\bibfnamefont {S.}~\bibnamefont {Klyatskaya}}, \bibinfo {author}
		{\bibfnamefont {M.}~\bibnamefont {Ruben}},\ and\ \bibinfo {author}
		{\bibfnamefont {W.}~\bibnamefont {Wernsdorfer}},\ }\bibfield  {title}
	{\enquote {\bibinfo {title} {{Electrically driven nuclear spin resonance in
					single-molecule magnets}},}\ }\href@noop {} {\bibfield  {journal} {\bibinfo
			{journal} {Science}\ }\textbf {\bibinfo {volume} {344}},\ \bibinfo {pages}
		{1135--1138} (\bibinfo {year} {2014})}\BibitemShut {NoStop}%
	\bibitem [{\citenamefont {Atzori}\ and\ \citenamefont
		{Sessoli}(2019)}]{Atzori2019}%
	\BibitemOpen
	\bibfield  {author} {\bibinfo {author} {\bibfnamefont {M.}~\bibnamefont
			{Atzori}}\ and\ \bibinfo {author} {\bibfnamefont {R.}~\bibnamefont
			{Sessoli}},\ }\bibfield  {title} {\enquote {\bibinfo {title} {{The Second
					Quantum Revolution: Role and Challenges of Molecular Chemistry}},}\
	}\href@noop {} {\bibfield  {journal} {\bibinfo  {journal} {J. Am. Chem.
				Soc.}\ }\textbf {\bibinfo {volume} {141}},\ \bibinfo {pages} {11339--11352}
		(\bibinfo {year} {2019})}\BibitemShut {NoStop}%
	\bibitem [{\citenamefont {Coronado}(2020)}]{Coronado2020}%
	\BibitemOpen
	\bibfield  {author} {\bibinfo {author} {\bibfnamefont {E.}~\bibnamefont
			{Coronado}},\ }\bibfield  {title} {\enquote {\bibinfo {title} {{Molecular
					magnetism: from chemical design to spin control in molecules, materials and
					devices}},}\ }\href@noop {} {\bibfield  {journal} {\bibinfo  {journal} {Nat.
				Rev. Mater.}\ }\textbf {\bibinfo {volume} {5}},\ \bibinfo {pages} {87--104}
		(\bibinfo {year} {2020})}\BibitemShut {NoStop}%
	\bibitem [{\citenamefont {Schuster}\ \emph {et~al.}(2005)\citenamefont
		{Schuster}, \citenamefont {Wallraff}, \citenamefont {Blais}, \citenamefont
		{Frunzio}, \citenamefont {Huang}, \citenamefont {Majer}, \citenamefont
		{Girvin},\ and\ \citenamefont {Schoelkopf}}]{Schuster2005}%
	\BibitemOpen
	\bibfield  {author} {\bibinfo {author} {\bibfnamefont {D.~I.}\ \bibnamefont
			{Schuster}}, \bibinfo {author} {\bibfnamefont {A.}~\bibnamefont {Wallraff}},
		\bibinfo {author} {\bibfnamefont {A.}~\bibnamefont {Blais}}, \bibinfo
		{author} {\bibfnamefont {L.}~\bibnamefont {Frunzio}}, \bibinfo {author}
		{\bibfnamefont {R.~S.}\ \bibnamefont {Huang}}, \bibinfo {author}
		{\bibfnamefont {J.}~\bibnamefont {Majer}}, \bibinfo {author} {\bibfnamefont
			{S.~M.}\ \bibnamefont {Girvin}},\ and\ \bibinfo {author} {\bibfnamefont
			{R.~J.}\ \bibnamefont {Schoelkopf}},\ }\bibfield  {title} {\enquote {\bibinfo
			{title} {{ac Stark shift and dephasing of a superconducting qubit strongly
					coupled to a cavity field}},}\ }\href@noop {} {\bibfield  {journal} {\bibinfo
			{journal} {Phys. Rev. Lett.}\ }\textbf {\bibinfo {volume} {94}},\ \bibinfo
		{pages} {123602} (\bibinfo {year} {2005})}\BibitemShut {NoStop}%
	\bibitem [{\citenamefont {Wallraff}\ \emph {et~al.}(2005)\citenamefont
		{Wallraff}, \citenamefont {Schuster}, \citenamefont {Blais}, \citenamefont
		{Frunzio}, \citenamefont {Majer}, \citenamefont {Devoret}, \citenamefont
		{Girvin},\ and\ \citenamefont {Schoelkopf}}]{Wallraff2005}%
	\BibitemOpen
	\bibfield  {author} {\bibinfo {author} {\bibfnamefont {A.}~\bibnamefont
			{Wallraff}}, \bibinfo {author} {\bibfnamefont {D.~I.}\ \bibnamefont
			{Schuster}}, \bibinfo {author} {\bibfnamefont {A.}~\bibnamefont {Blais}},
		\bibinfo {author} {\bibfnamefont {L.}~\bibnamefont {Frunzio}}, \bibinfo
		{author} {\bibfnamefont {J.}~\bibnamefont {Majer}}, \bibinfo {author}
		{\bibfnamefont {M.~H.}\ \bibnamefont {Devoret}}, \bibinfo {author}
		{\bibfnamefont {S.~M.}\ \bibnamefont {Girvin}},\ and\ \bibinfo {author}
		{\bibfnamefont {R.~J.}\ \bibnamefont {Schoelkopf}},\ }\bibfield  {title}
	{\enquote {\bibinfo {title} {{Approaching unit visibility for control of a
					superconducting qubit with dispersive readout}},}\ }\href@noop {} {\bibfield
		{journal} {\bibinfo  {journal} {Phys. Rev. Lett.}\ }\textbf {\bibinfo
			{volume} {95}},\ \bibinfo {pages} {060501} (\bibinfo {year}
		{2005})}\BibitemShut {NoStop}%
	\bibitem [{\citenamefont {Sheludiakov}\ \emph {et~al.}(2019)\citenamefont
		{Sheludiakov}, \citenamefont {Ahokas}, \citenamefont {J{\"{a}}rvinen},
		\citenamefont {Lehtonen}, \citenamefont {Vasiliev}, \citenamefont {Dmitriev},
		\citenamefont {Lee},\ and\ \citenamefont {Khmelenko}}]{Sheludiakov2019}%
	\BibitemOpen
	\bibfield  {author} {\bibinfo {author} {\bibfnamefont {S.}~\bibnamefont
			{Sheludiakov}}, \bibinfo {author} {\bibfnamefont {J.}~\bibnamefont {Ahokas}},
		\bibinfo {author} {\bibfnamefont {J.}~\bibnamefont {J{\"{a}}rvinen}},
		\bibinfo {author} {\bibfnamefont {L.}~\bibnamefont {Lehtonen}}, \bibinfo
		{author} {\bibfnamefont {S.}~\bibnamefont {Vasiliev}}, \bibinfo {author}
		{\bibfnamefont {Y.~A.}\ \bibnamefont {Dmitriev}}, \bibinfo {author}
		{\bibfnamefont {D.~M.}\ \bibnamefont {Lee}},\ and\ \bibinfo {author}
		{\bibfnamefont {V.~V.}\ \bibnamefont {Khmelenko}},\ }\bibfield  {title}
	{\enquote {\bibinfo {title} {{Electrons Trapped in Solid Neon–Hydrogen
					Mixtures Below 1K}},}\ }\href@noop {} {\bibfield  {journal} {\bibinfo
			{journal} {J. Low Temp. Phys.}\ }\textbf {\bibinfo {volume} {195}},\ \bibinfo
		{pages} {365--377} (\bibinfo {year} {2019})}\BibitemShut {NoStop}%
	\bibitem [{\citenamefont {Jacobsen}, \citenamefont {Penoncello},\ and\
		\citenamefont {Lemmon}(1997)}]{jacobsen1997thermodynamic}%
	\BibitemOpen
	\bibfield  {author} {\bibinfo {author} {\bibfnamefont {R.~T.}\ \bibnamefont
			{Jacobsen}}, \bibinfo {author} {\bibfnamefont {S.~G.}\ \bibnamefont
			{Penoncello}},\ and\ \bibinfo {author} {\bibfnamefont {E.~W.}\ \bibnamefont
			{Lemmon}},\ }\bibfield  {title} {\enquote {\bibinfo {title} {Thermodynamic
				properties of cryogenic fluids},}\ }in\ \href@noop {} {\emph {\bibinfo
			{booktitle} {Thermodynamic Properties of Cryogenic Fluids}}}\ (\bibinfo
	{publisher} {Springer},\ \bibinfo {year} {1997})\ pp.\ \bibinfo {pages}
	{31--287}\BibitemShut {NoStop}%
	\bibitem [{\citenamefont {Pollack}(1964)}]{Pollack1964}%
	\BibitemOpen
	\bibfield  {author} {\bibinfo {author} {\bibfnamefont {G.~L.}\ \bibnamefont
			{Pollack}},\ }\bibfield  {title} {\enquote {\bibinfo {title} {{The Solid
					State of Rare Gases}},}\ }\href@noop {} {\bibfield  {journal} {\bibinfo
			{journal} {Rev. Mod. Phys.}\ }\textbf {\bibinfo {volume} {36}},\ \bibinfo
		{pages} {748} (\bibinfo {year} {1964})}\BibitemShut {NoStop}%
	\bibitem [{\citenamefont {Batchelder}, \citenamefont {Losee},\ and\
		\citenamefont {Simmons}(1967)}]{Batchelder1967}%
	\BibitemOpen
	\bibfield  {author} {\bibinfo {author} {\bibfnamefont {D.~N.}\ \bibnamefont
			{Batchelder}}, \bibinfo {author} {\bibfnamefont {D.~L.}\ \bibnamefont
			{Losee}},\ and\ \bibinfo {author} {\bibfnamefont {R.~O.}\ \bibnamefont
			{Simmons}},\ }\bibfield  {title} {\enquote {\bibinfo {title} {{Measurements
					of lattice constant, thermal expansion, and isothermal compressibility of
					neon single crystals}},}\ }\href@noop {} {\bibfield  {journal} {\bibinfo
			{journal} {Phys. Rev.}\ }\textbf {\bibinfo {volume} {162}},\ \bibinfo {pages}
		{767} (\bibinfo {year} {1967})}\BibitemShut {NoStop}%
	\bibitem [{\citenamefont {Zavyalov}\ \emph {et~al.}(2005)\citenamefont
		{Zavyalov}, \citenamefont {Smolyaninov}, \citenamefont {Zotova},
		\citenamefont {Borodin},\ and\ \citenamefont
		{Bogomolov}}]{zavyalov2005electron}%
	\BibitemOpen
	\bibfield  {author} {\bibinfo {author} {\bibfnamefont {V.}~\bibnamefont
			{Zavyalov}}, \bibinfo {author} {\bibfnamefont {I.}~\bibnamefont
			{Smolyaninov}}, \bibinfo {author} {\bibfnamefont {E.}~\bibnamefont {Zotova}},
		\bibinfo {author} {\bibfnamefont {A.}~\bibnamefont {Borodin}},\ and\ \bibinfo
		{author} {\bibfnamefont {S.}~\bibnamefont {Bogomolov}},\ }\bibfield  {title}
	{\enquote {\bibinfo {title} {Electron states above the surfaces of solid
				cryodielectrics for quantum-computing.}}\ }\href@noop {} {\bibfield
		{journal} {\bibinfo  {journal} {J. Low Temp. Phys.}\ }\textbf {\bibinfo
			{volume} {138}},\ \bibinfo {pages} {415--420} (\bibinfo {year}
		{2005})}\BibitemShut {NoStop}%
	\bibitem [{\citenamefont {Leiderer}, \citenamefont {Kono},\ and\ \citenamefont
		{Rees}(2016)}]{leiderer2016cryocrystals}%
	\BibitemOpen
	\bibfield  {author} {\bibinfo {author} {\bibfnamefont {P.}~\bibnamefont
			{Leiderer}}, \bibinfo {author} {\bibfnamefont {K.}~\bibnamefont {Kono}},\
		and\ \bibinfo {author} {\bibfnamefont {D.}~\bibnamefont {Rees}},\ }\bibfield
	{title} {\enquote {\bibinfo {title} {Cryocrystals as substrates for surface
				state electrons},}\ }in\ \href@noop {} {\emph {\bibinfo {booktitle} {The 11th
				International Conference on Cryocrystals and Quantum Crystals}}}\ (\bibinfo
	{year} {2016})\ pp.\ \bibinfo {pages} {67--67}\BibitemShut {NoStop}%
	\bibitem [{\citenamefont {Kajita}(1984)}]{Kajita1984}%
	\BibitemOpen
	\bibfield  {author} {\bibinfo {author} {\bibfnamefont {K.}~\bibnamefont
			{Kajita}},\ }\bibfield  {title} {\enquote {\bibinfo {title} {{A new
					two-dimensional electron system on the surface of solid neon}},}\ }\href@noop
	{} {\bibfield  {journal} {\bibinfo  {journal} {Surf. Sci.}\ }\textbf
		{\bibinfo {volume} {142}},\ \bibinfo {pages} {86--95} (\bibinfo {year}
		{1984})}\BibitemShut {NoStop}%
	\bibitem [{\citenamefont {Nilsson}, \citenamefont {Pettersson},\ and\
		\citenamefont {Norskov}(2011)}]{nilsson2011chemical}%
	\BibitemOpen
	\bibfield  {author} {\bibinfo {author} {\bibfnamefont {A.}~\bibnamefont
			{Nilsson}}, \bibinfo {author} {\bibfnamefont {L.~G.}\ \bibnamefont
			{Pettersson}},\ and\ \bibinfo {author} {\bibfnamefont {J.}~\bibnamefont
			{Norskov}},\ }\href@noop {} {\emph {\bibinfo {title} {Chemical bonding at
				surfaces and interfaces}}}\ (\bibinfo  {publisher} {Elsevier},\ \bibinfo
	{year} {2011})\BibitemShut {NoStop}%
	\bibitem [{\citenamefont {Ibach}(2006)}]{ibach2006physics}%
	\BibitemOpen
	\bibfield  {author} {\bibinfo {author} {\bibfnamefont {H.}~\bibnamefont
			{Ibach}},\ }\href@noop {} {\emph {\bibinfo {title} {Physics of surfaces and
				interfaces}}},\ Vol.\ \bibinfo {volume} {2006}\ (\bibinfo  {publisher}
	{Springer},\ \bibinfo {year} {2006})\BibitemShut {NoStop}%
	\bibitem [{\citenamefont {Pozar}(2011)}]{Pozar2011}%
	\BibitemOpen
	\bibfield  {author} {\bibinfo {author} {\bibfnamefont {D.~M.}\ \bibnamefont
			{Pozar}},\ }\href@noop {} {\emph {\bibinfo {title} {{Microwave
					Engineering}}}}\ (\bibinfo  {publisher} {Wiley},\ \bibinfo {year}
	{2011})\BibitemShut {NoStop}%
	\bibitem [{\citenamefont {Walls}\ and\ \citenamefont
		{Milburn}(2007)}]{walls2007quantum}%
	\BibitemOpen
	\bibfield  {author} {\bibinfo {author} {\bibfnamefont {D.~F.}\ \bibnamefont
			{Walls}}\ and\ \bibinfo {author} {\bibfnamefont {G.~J.}\ \bibnamefont
			{Milburn}},\ }\href@noop {} {\emph {\bibinfo {title} {Quantum optics}}}\
	(\bibinfo  {publisher} {Springer Science \& Business Media},\ \bibinfo {year}
	{2007})\BibitemShut {NoStop}%
	\bibitem [{\citenamefont {{D. I. Schuster}}(2007)}]{DavidIsaacSchuster2007}%
	\BibitemOpen
	\bibfield  {author} {\bibinfo {author} {\bibnamefont {{D. I. Schuster}}},\
	}\emph {\bibinfo {title} {{Circuit Quantum Electrodynamics}}},\ \href
	{http://www.eng.yale.edu/rslab/papers/SchusterThesis.pdf{\%}5Cnpapers://b942896d-313a-49f0-995e-1484170e0fc9/Paper/p257}
	{Ph.D. thesis},\ \bibinfo  {school} {Yale University} (\bibinfo {year}
	{2007})\BibitemShut {NoStop}%
	\bibitem [{\citenamefont {Krantz}\ \emph {et~al.}(2019)\citenamefont {Krantz},
		\citenamefont {Kjaergaard}, \citenamefont {Yan}, \citenamefont {Orlando},
		\citenamefont {Gustavsson},\ and\ \citenamefont {Oliver}}]{Krantz2019}%
	\BibitemOpen
	\bibfield  {author} {\bibinfo {author} {\bibfnamefont {P.}~\bibnamefont
			{Krantz}}, \bibinfo {author} {\bibfnamefont {M.}~\bibnamefont {Kjaergaard}},
		\bibinfo {author} {\bibfnamefont {F.}~\bibnamefont {Yan}}, \bibinfo {author}
		{\bibfnamefont {T.~P.}\ \bibnamefont {Orlando}}, \bibinfo {author}
		{\bibfnamefont {S.}~\bibnamefont {Gustavsson}},\ and\ \bibinfo {author}
		{\bibfnamefont {W.~D.}\ \bibnamefont {Oliver}},\ }\bibfield  {title}
	{\enquote {\bibinfo {title} {{A quantum engineer's guide to superconducting
					qubits}},}\ }\href@noop {} {\bibfield  {journal} {\bibinfo  {journal} {Appl.
				Phys. Rev.}\ }\textbf {\bibinfo {volume} {6}},\ \bibinfo {pages} {021318}
		(\bibinfo {year} {2019})}\BibitemShut {NoStop}%
	\bibitem [{\citenamefont {Ithier}\ \emph {et~al.}(2005)\citenamefont {Ithier},
		\citenamefont {Collin}, \citenamefont {Joyez}, \citenamefont {Meeson},
		\citenamefont {Vion}, \citenamefont {Esteve}, \citenamefont {Chiarello},
		\citenamefont {Shnirman}, \citenamefont {Makhlin}, \citenamefont {Schriefl},\
		and\ \citenamefont {Sch\"on}}]{Ithier2005}%
	\BibitemOpen
	\bibfield  {author} {\bibinfo {author} {\bibfnamefont {G.}~\bibnamefont
			{Ithier}}, \bibinfo {author} {\bibfnamefont {E.}~\bibnamefont {Collin}},
		\bibinfo {author} {\bibfnamefont {P.}~\bibnamefont {Joyez}}, \bibinfo
		{author} {\bibfnamefont {P.~J.}\ \bibnamefont {Meeson}}, \bibinfo {author}
		{\bibfnamefont {D.}~\bibnamefont {Vion}}, \bibinfo {author} {\bibfnamefont
			{D.}~\bibnamefont {Esteve}}, \bibinfo {author} {\bibfnamefont
			{F.}~\bibnamefont {Chiarello}}, \bibinfo {author} {\bibfnamefont
			{A.}~\bibnamefont {Shnirman}}, \bibinfo {author} {\bibfnamefont
			{Y.}~\bibnamefont {Makhlin}}, \bibinfo {author} {\bibfnamefont
			{J.}~\bibnamefont {Schriefl}},\ and\ \bibinfo {author} {\bibfnamefont
			{G.}~\bibnamefont {Sch\"on}},\ }\bibfield  {title} {\enquote {\bibinfo
			{title} {Decoherence in a superconducting quantum bit circuit},}\ }\href@noop
	{} {\bibfield  {journal} {\bibinfo  {journal} {Phys. Rev. B}\ }\textbf
		{\bibinfo {volume} {72}},\ \bibinfo {pages} {134519} (\bibinfo {year}
		{2005})}\BibitemShut {NoStop}%
	\bibitem [{\citenamefont {Chen}(2018)}]{Chen2018}%
	\BibitemOpen
	\bibfield  {author} {\bibinfo {author} {\bibfnamefont {Z.}~\bibnamefont
			{Chen}},\ }\emph {\bibinfo {title} {Metrology of Quantum Control and
			Measurement in Superconducting Qubits}},\ \href@noop {} {Ph.D. thesis},\
	\bibinfo  {school} {University of California Santa Barbara} (\bibinfo {year}
	{2018})\BibitemShut {NoStop}%
\end{thebibliography}
\end{document}